\documentclass[12pt]{article}
\usepackage{amssymb,bm,epsfig}

\renewcommand{\thefootnote}{\fnsymbol{footnote}}
\newcommand{\vev}[1]{{\langle{#1}\rangle}}

\begin{document}

\title{
\begin{flushright}
\begin{minipage}{0.2\linewidth}
\normalsize
WU-HEP-12-06 \\
KUNS-2421 \\*[50pt]
\end{minipage}
\end{flushright}
{\Large \bf 
Phenomenological aspects of 10D SYM theory 
with magnetized extra dimensions
\\*[20pt]}}

\author{Hiroyuki~Abe$^{1,}$\footnote{
E-mail address: abe@waseda.jp}, \ 
Tatsuo~Kobayashi$^{2,}$\footnote{
E-mail address: kobayash@gauge.scphys.kyoto-u.ac.jp}, \ 
Hiroshi~Ohki$^{3,}$\footnote{
E-mail address: ohki@kmi.nagoya-u.ac.jp}, \\ 
Akane~Oikawa$^{1,}$\footnote{
E-mail address: cosmic-space.46y@ruri.waseda.jp}
\ and \ 
Keigo~Sumita$^{1,}$\footnote{
E-mail address: k.sumita@moegi.waseda.jp
}\\*[20pt]
$^1${\it \normalsize 
Department of Physics, Waseda University, 
Tokyo 169-8555, Japan} \\
$^2${\it \normalsize 
Department of Physics, Kyoto University, 
Kyoto 606-8502, Japan} \\
$^3${\it \normalsize 
Kobayashi-Maskawa Institute for the 
Origin of Particles and the Universe (KMI),} \\
{\it \normalsize 
Nagoya University, Nagoya 464-8602, Japan} \\*[50pt]}

\date{
\centerline{\small \bf Abstract}
\begin{minipage}{0.9\linewidth}
\medskip 
\medskip 
\small
We present a particle physics model based on a ten-dimensional (10D)
 super Yang-Mills (SYM) theory compactified on magnetized 
tori preserving four-dimensional ${\cal N}=1$ supersymmetry. 
The low-energy spectrum contains the minimal supersymmetric 
standard model with hierarchical Yukawa couplings caused by a 
wavefunction localization of the chiral matter fields due to 
the existence of magnetic fluxes, allowing a semi-realistic 
pattern of the quark and the lepton masses and mixings. 
We show supersymmetric flavor structures at low energies 
induced by a moduli-mediated and an anomaly-mediated 
supersymmetry breaking. 
\end{minipage}
}

\begin{titlepage}
\maketitle
\thispagestyle{empty}
\clearpage
\tableofcontents
\thispagestyle{empty}
\end{titlepage}

\renewcommand{\thefootnote}{\arabic{footnote}}
\setcounter{footnote}{0}

\section{Introduction}

The standard model (SM) of elementary particles is a quite 
successful theory, consistent with all the experimental 
data obtained so far with a great accuracy. There are, 
however, many free parameters, which can not be determined 
theoretically, making the model less predictable. 
Among these parameters, especially, Yukawa coupling constants 
seem to be awfully hierarchical in order to explain the observed 
masses and mixing angles of the quarks and the leptons. It is argued 
that some flavor symmetries are helpful to understand such a 
hierarchical structure. 
(See, for a review, Ref.~\cite{Ishimori:2010au}.) 
Another interesting possibility is a quasi-localization of 
matter fields in extra dimensions, where the hierarchical 
couplings are obtained from the overlap integral of their 
localized wavefunctions~\cite{ArkaniHamed:1999dc}.
It is also suggested that the former flavor symmetries are 
realized geometrically as a consequence of the latter 
wavefunction localization in extra 
dimensions~\cite{Abe:2009vi,Abe:2010iv}
\footnote{Non-Abelian discrete flavor symmetries are also 
obtained within the framework of heterotic string theory on orbifolds
\cite{Kobayashi:2004ya}.}.

The SM does not describe gravitational interactions of 
elementary particles that could play an important role 
at the very beginning of our universe. 
Superstring theories in ten-dimensional (10D) spacetime are almost 
the only known candidates that can treat gravitational interactions 
at the quantum level. These theories possess few free parameters 
and potentially more predictive than the SM. 
Supersymmetric Yang-Mills (SYM) theories in various spacetime 
dimensions appear as low energy effective theories of superstring 
compactifications with or without D-branes. 
Thus, it is an interesting possibility that the SM is embedded in 
one of such SYM theories, that is, the SM is realized as a low 
energy effective theory of the superstrings. 
In such a string model building, how to break higher-dimensional 
supersymmetry and to obtain a chiral spectrum is the key issue. 
String compactifications on the Calabi-Yau (CY) space provide a 
general procedure for such a purpose. However, the metric of 
a generic CY space is hard to be determined analytically, 
that makes the phenomenological studies qualitative, 
but not quantitative. 

It is quite interesting that even simple toroidal compactifications 
but with magnetic fluxes in extra dimensions induce chiral 
spectra~\cite{Angelantonj:2000hi,Cremades:2004wa} in higher-dimensional SYM theories. 
The higher-dimensional supersymmetry such as ${\cal N}=4$ in terms 
of supercharges in four-dimensional (4D) spacetime is broken by the 
magnetic fluxes down to 4D ${\cal N}=0$, $1$ or $2$ depending on the 
configuration of fluxes. The number of the chiral zero-modes is 
determined by the number of magnetic fluxes. A phenomenologically 
attractive feature is that these chiral zero modes localize toward 
different points in magnetized extra dimensions. 
The overlap integrals of localized wavefunctions yield hierarchical 
couplings in the 4D effective theory of these zero modes, that 
could explain, e.g., observed hierarchical masses and mixing angles 
of the quarks and the leptons~\cite{Abe:2008sx}. 
Furthermore, higher-order couplings can also be computed as 
the overlap integrals of wavefunctions~\cite{Abe:2009dr}.
A theoretically attractive point here is that many peculiar 
properties of the SM, such as the 4D chirality, the number 
of generations, the flavor symmetries~\cite{Abe:2009vi,Abe:2009uz,BerasaluceGonzalez:2012vb} 
and potentially hierarchical Yukawa couplings 
all could be determined by the magnetic fluxes. 

Moreover if the 4D ${\cal N}=1$ supersymmetry remains, 
a supersymmetric standard model could be realized below the 
compactification scale that has many attractive features 
beyond the SM, such as the lightest supersymmetric particle 
as a candidate of dark matter and so on. 
In our previous work~\cite{Abe:2012ya}, we have presented 
4D ${\cal N}=1$ superfield description of 10D SYM theories 
compactified on magnetized tori which preserve the 
${\cal N}=1$ supersymmetry, and derived 4D effective action 
for massless zero-modes written in the ${\cal N}=1$ superspace. 
We further identified moduli dependence of the effective 
action by promoting the Yang-Mills (YM) gauge coupling constant 
$g$ and geometric parameters $R_i$ and $\tau_i$ to a dilaton, 
K\"ahler and complex-structure moduli superfields, which 
allows an explicit estimation of soft supersymmetry breaking 
parameters in the supersymmetric SM caused by moduli-mediated 
supersymmetry breaking. 
The resulting effective supergravity action would be useful 
for building phenomenological models and for analyzing them 
systematically. 

Motivated by the above arguments, in this paper, we construct 
a particle physics model based on 10D SYM theory compactified 
on three factorizable tori $T^2 \times T^2 \times T^2$ where 
magnetic fluxes are present in the YM sector. 
We search a phenomenologically viable flux configuration that 
induces a 4D chiral spectrum including the minimal supersymmetric 
standard model (MSSM), based on the effective action written in 
${\cal N}=1$ superspace. For such a flux configuration that realize 
a realistic pattern of the quark and the lepton masses and their mixing angles, 
we further estimate the sizes of supersymmetric flavor violations 
caused by the moduli-mediated supersymmetry breaking. 

The sections are organized as follows. 
In Sec.~\ref{sec:10d}, a superfield description of the 10D SYM 
theory is briefly reviewed based on Ref.~\cite{Abe:2012ya}, 
which allows the systematic introduction of magnetic fluxes 
in extra dimensions preserving the ${\cal N}=1$ supersymmetry. 
Then, we construct a model that contains the spectrum of the MSSM, 
in which the most massless exotic modes are projected out 
due to the existence of the magnetic fluxes and a certain orbifold 
projection in Sec.~\ref{sec:model}. 
In Sec.~\ref{sec:flavor}, we numerically search a location in the 
moduli space of the model where a realistic pattern of the quark 
and the lepton masses and their mixing angles are obtained. Then, assuming the 
moduli-mediated supersymmetry breaking, we estimate the magnitude 
of the mass insertion parameters representing typical sizes of 
various flavor changing neutral currents (FCNC) in Sec.~\ref{sec:pheno}. 
Sec.~\ref{sec:conc} is devoted to conclusions and discussions. 
In Appendix~\ref{app:kmhy}, the K\"ahler metrics and the holomorphic 
Yukawa couplings are exhibited for the MSSM matter fields 
in the 4D effective theory.

\section{The 10D SYM theory in ${\cal N}=1$ superspace}
\label{sec:10d}

Based on Ref.~\cite{Abe:2012ya}, 
in this section, we review a compactification of 10D SYM 
theory on 4D flat  Minkowski spacetime times a product of 
factorizable three tori $T^2 \times T^2 \times T^2$ 
and a superfield description suitable for such a 
compactification with magnetic fluxes in each torus 
preserving 4D ${\cal N}=1$ supersymmetry. 
The geometric (torus) parameter dependence is 
explicitly shown in this procedure, which is important 
to determine couplings between YM and moduli superfields 
in the 4D effective action for chiral zero-modes. 

The 10D SYM theory is described by the following action, 
\begin{eqnarray}
S &=& \int d^{10}X\,\sqrt{-G}\,
\frac{1}{g^2}\,{\rm Tr}\left[ 
-\frac{1}{4}F^{MN}F_{MN} 
+\frac{i}{2} \bar\lambda \Gamma^M D_M \lambda 
\right], 
\label{eq:10dsym}
\end{eqnarray}
where $g$ is a 10D YM gauge coupling constant 
and the trace is performed over the adjoint representation 
of the YM gauge group. The 10D spacetime coordinates are 
denoted by $X^M$, and the vector/tensor indices 
$M,N=0,1,\ldots,9$ are lowered and raised by the 10D 
metric $G_{MN}$ and its inverse $G^{MN}$, respectively. 
The YM field strength $F_{MN}$ and 
the covariant derivative $D_M$ are given by 
$F_{MN} = 
\partial_M A_N - \partial_N A_M -i [A_M,A_N]$ and 
$D_M \lambda = \partial_M \lambda -i [A_M,\lambda]$ 
for a 10D vector (gauge) field $A_M$ and 
a 10D Majorana-Weyl spinor field $\lambda$, respectively. 
The spinor field $\lambda$ satisfies 
10D Majorana and Weyl conditions, 
$\lambda^C = \lambda$ and 
$\Gamma \lambda = +\lambda$, respectively, 
where $\lambda^C$ denotes a 10D charge conjugation of 
$\lambda$, and $\Gamma$ is a 10D chirality operator. 

The 10D spacetime (real) coordinates $X^M=(x^\mu,y^m)$ 
are decomposed into 4D Minkowski spacetime coordinates 
$x^\mu$ with $\mu=0,1,2,3$ and six dimensional (6D) 
extra space coordinates $y^m$ with $m=4,\ldots,9$.  
The zeroth component $\mu=0$ describes the time component. 
The 10D vector field is similarly decomposed as 
$A_M=(A_\mu,A_m)$. 
The 10D background metric is given by 
\begin{eqnarray}
ds^2 &=& G_{NN}dX^MdX^N \ = \ 
\eta_{\mu \nu} dx^\mu dx^\nu + g_{mn} dy^m dy^n, 
\nonumber
\end{eqnarray}
where 
$\eta_{\mu \nu}={\rm diag}(-1,+1,+1,+1)$. 
Because we consider a torus compactification of internal 
6D space $y^m$ by identifying $y^m \sim y^m + 2$ and 
the 6D torus is decomposed as a product of factorizable 
three tori, $T^2 \times T^2 \times T^2$, 
the extra 6D metric can be described as 
\begin{eqnarray}
g_{mn} &=& \left( 
\begin{array}{ccc}
g^{(1)} & 0 & 0 \\
0 & g^{(2)}& 0 \\
0 & 0 & g^{(3)} 
\end{array} \right), 
\nonumber
\end{eqnarray}
where each entry is a $2 \times 2$ matrix and 
the diagonal submatrices are expressed as 
\begin{eqnarray}
g^{(i)} &=& (2 \pi R_i)^2 \left( 
\begin{array}{cc}
1 & {\rm Re}\,\tau_i \\
{\rm Re}\,\tau_i & |\tau_i|^2
\end{array} \right), 
\nonumber
\end{eqnarray}
for $i=1,2,3$. 
The real and the complex parameters $R_i$ and $\tau_i$ 
determine the size and the shape of the 
$i$th torus $T^2$, respectively. The area ${\cal A}^{(i)}$ of the $i$th 
torus is determined by these parameters as 
\begin{eqnarray}
{\cal A}^{(i)} &=& (2 \pi R_i)^2\,{\rm Im}\, \tau_i. 
\nonumber
\end{eqnarray}

The complex coordinates $z^i$ for $i=1,2,3$ defined by 
\begin{eqnarray}
z^i &\equiv& \frac{1}{2}(y^{2+2i} + \tau_i\,y^{3+2i}), 
\qquad 
\bar{z}^{\bar{i}} \ \equiv \ (z^i)^\ast, 
\nonumber
\end{eqnarray}
are extremely useful for describing the action in 
4D ${\cal N}=1$ superspace, where the corresponding 
complex vector components $A_i$ are defined by 
\begin{eqnarray}
A_i &\equiv& 
-\frac{1}{{\rm Im}\,\tau_i} 
(\tau_i^\ast \,A_{2+2i}-A_{3+2i}), 
\qquad 
\bar{A}_{\bar{i}} \ \equiv \ (A_i)^\dagger. 
\nonumber
\end{eqnarray}
In the complex coordinate, 
the torus boundary conditions are expressed as 
$z^i \sim z^i +1$ and $z^i \sim  z^i +\tau^i$, 
and the metric is found as 
$h_{i\bar{j}} = 
2\,(2 \pi R_i)^2\,\delta_{i\bar{j}} 
\ = \ \delta_{{\rm i}\bar{\rm j}}\,
e_i^{\ {\rm i}}\,\bar{e}_{\bar{j}}^{\ \bar{\rm j}}$ 
satisfying 
$2h_{i\bar{j}}dz^id\bar{z}^{\bar{j}} 
= g_{mn}dy^mdy^n = ds_{6D}^2$, 
where 
$e_i^{\ {\rm i}}=
\sqrt{2}\,(2 \pi R_i)\,\delta_i^{\ {\rm i}}$ 
is a vielbein, and 
the Roman indices represent local Lorentz space. 
The Italic (Roman) indices $i,j,\ldots$ 
(${\rm i}$, ${\rm j}, \ldots$) are 
lowered and raised by the metric $h_{i\bar{j}}$ 
and its inverse $h^{\bar{i}j}$ 
($\delta_{{\rm i}\bar{\rm j}}$ 
and its inverse $\delta^{\bar{\rm i}{\rm j}}$), 
respectively. 

The 10D SYM theory possesses ${\cal N}=4$ supersymmetry 
counted by a 4D supercharge. The YM vector and spinor 
fields, $A_M$ and $\lambda$, are decomposed into (on-shell) 
4D ${\cal N}=1$ single vector and triple chiral multiplets, 
${\bm V}=\left\{ A_\mu, \lambda_0 \right\}$ and 
${\bm \phi}_i \ = \ \left\{ A_i, \lambda_i \right\}$ $(i=1,2,3)$,
respectively, where the 10D Majorana-Weyl spinor $\lambda$ 
is decomposed into four 4D Weyl (or equivalently Majorana) 
spinors $\lambda_0$ and $\lambda_i$. 
If we write the chirality associated with 6D spacetime 
coordinates $(x^\mu,z^i)$ in the $i$th subscript of 
$\lambda$ like $\lambda_{\pm \pm \pm}$, 
the decomposed spinor fields $\lambda_0$ and $\lambda_i$ 
are identified with the chirality eigenstates 
$\lambda_{\pm \pm \pm}$ as 
$\lambda_0 = \lambda_{+++}$, 
$\lambda_1 = \lambda_{+--}$, 
$\lambda_2 = \lambda_{-+-}$ and 
$\lambda_3 = \lambda_{--+}$ 
for the 4D chirality fixed, e.g. the positive chirality. 
Note that the components 
$\lambda_{---}$, $\lambda_{-++}$, 
$\lambda_{+-+}$ and $\lambda_{++-}$ 
do not exist in the 10D Majorana-Weyl spinor $\lambda$ 
due to the condition $\Gamma \lambda = +\lambda$. 

The above ${\cal N}=1$ vector and chiral multiplets, 
${\bm V}$ and ${\bm \phi}_i$, are expressed by vector and 
chiral superfields, $V$ and $\phi_i$, respectively as 
\begin{eqnarray}
V &\equiv& -\theta\sigma^\mu\bar\theta A_\mu
+i\bar\theta\bar\theta\theta\lambda_0 
-i\theta\theta\bar\theta\bar\lambda_0 
+\frac12\theta\theta\bar\theta\bar\theta D, 
\nonumber \\
\phi_i &\equiv& \frac1{\sqrt2} A_i
+\sqrt2\theta\lambda_i+\theta\theta F_i, 
\nonumber
\end{eqnarray}
where $\theta$ and $\bar\theta$ are 
Grassmann coordinates of 4D ${\cal N}=1$ superspace. 
The 10D SYM action (\ref{eq:10dsym}) can be written in the 
${\cal N}=1$ superspace as~\cite{Marcus:1983wb} 
\begin{eqnarray}
S &=& \int d^{10}X \sqrt{-G} \left[ \int d^4\theta\,{\cal K}
+\left\{ \int d^2 \theta\,\left( 
\frac{1}{4g^2} {\cal W}^\alpha {\cal W}_\alpha 
+ {\cal W} \right) +{\rm h.c.} \right\} 
\right]. 
\nonumber
\end{eqnarray}
The functions of the superfields, 
${\cal K}$, ${\cal W}$ and ${\cal W}_\alpha$, are given by 
\begin{eqnarray}
{\cal K} &=& \frac{2}{g^2}  h^{\bar{i}j} 
{\rm Tr} \left[\left(\sqrt2\bar\partial_{\bar{i}}
+\bar\phi_{\bar{i}}\right)e^{-V}
\left(-\sqrt2\partial_j + \phi_j \right)e^V
+\bar\partial_{\bar{i}} e^{-V} \partial_j e^V \right] 
+{\cal K}_{\rm WZW}, 
\nonumber \\
{\cal W} &=& \frac{1}{g^2} 
\epsilon^{{\rm i}{\rm j}{\rm k}} 
e_{{\rm i}}^{\ i} e_{{\rm j}}^{\ j} e_{{\rm k}}^{\ k} 
{\rm Tr} \left[ \sqrt{2}\, \phi_i
\left(\partial_j\phi_k-\frac{1}{3\sqrt2}
\left[\phi_j,\phi_k\right]\right)\right], 
\nonumber \\
{\cal W}_\alpha &=& 
-\frac{1}{4} \bar{D} \bar{D} e^{-V} D_\alpha e^V, 
\nonumber
\end{eqnarray}
where 
$\epsilon^{{\rm i}{\rm j}{\rm k}}$ 
is a totally antisymmetric tensor satisfying 
$\epsilon^{123}=1$, and 
$D_\alpha$ ($\bar{D}_{\dot\alpha}$) 
is a supercovariant derivative (its conjugate) 
with a 4D spinor index $\alpha$ ($\dot\alpha$). 
The term ${\cal K}_{\rm WZW}$ represents a 
Wess-Zumino-Witten term which vanishes 
in the Wess-Zumino (WZ) gauge. 

The equations of motion for 
auxiliary fields $D$ and $F_i$ lead to 
\begin{eqnarray}
D &=& - h^{\bar{i}j} \left( 
\bar\partial_{\bar{i}} A_j + \partial_j \bar{A}_{\bar{i}} 
+\frac{1}{2} \left[ \bar{A}_{\bar{i}}, A_j \right] \right), 
\label{eq:osd} \\
\bar{F}_{\bar{i}} &=& -h_{j\bar{i}}\, 
\epsilon^{{\rm j}{\rm k}{\rm l}} 
e_{{\rm j}}^{\ j} e_{{\rm k}}^{\ k} e_{{\rm l}}^{\ l} 
\left( \partial_k A_l 
-\frac{1}{4} \left[ A_k,\, A_l \right] \right). 
\label{eq:osfi}
\end{eqnarray}
The condition 
$\langle D \rangle = \langle F_i \rangle =0$ 
determines supersymmetric vacua. 
A trivial supersymmetric vacuum is given by 
$\langle A_i \rangle = 0$ where the full ${\cal N}=4$ 
supersymmetry as well as the YM gauge symmetry is preserved. 
In the following, we select one of nontrivial supersymmetric 
vacua where magnetic fluxes exist in the YM sector, 
and construct a particle physics model with a semi-realistic 
flavor structure of (s)quarks and (s)leptons caused by 
a wavefunction localization of chiral matter fields 
in extra dimensions due to the effect of magnetic fluxes.

\section{The model building}
\label{sec:model}

We consider the 10D $U(N)$ SYM theory\footnote{A similar study is 
possible by starting with other gauge groups \cite{Choi:2009pv}.}
 on a supersymmetric 
magnetic background where the YM fields take the following 
4D Lorentz invariant and at least ${\cal N}=1$ 
supersymmetric VEVs, 
\begin{eqnarray}
\vev{A_i} &=& \frac{\pi}{{\rm Im}\, \tau_i} 
\left( M^{(i)}\,\bar{z}_{\bar{i}}+\bar\zeta_i \right), \qquad 
\vev{A_\mu} \ = \ 
\vev{\lambda_0} \ = \ 
\vev{\lambda_i} \ = \ 
\vev{F_i} \ = \ 
\vev{D} \ = \ 0. 
\label{eq:ymvevs}
\end{eqnarray}
Here $N \times N$ diagonal matrices of Abelian magnetic fluxes 
and those of Wilson-lines are denoted, respectively, as 
\begin{eqnarray}
M^{(i)} &=& {\rm diag}
(M_1^{(i)},M_2^{(i)},\ldots,M_N^{(i)}), 
\label{eq:abelianflux} \\
\zeta_i &=& {\rm diag}
(\zeta_1^{(i)},\zeta_2^{(i)},\ldots,\zeta_N^{(i)}). 
\nonumber
\end{eqnarray}
The magnetic fluxes satisfying 
the Dirac's quantization condition, 
$M_1^{(i)},M_2^{(i)},\ldots,M_N^{(i)} \in {\bf Z}$, 
are further constrained by the supersymmetry conditions 
$\vev{D}=0$ and $\vev{F_i}=0$ in Eq.~(\ref{eq:ymvevs}), 
which are written as 
\begin{eqnarray}
h^{\bar{i}j} \left( \bar\partial_{\bar{i}} \vev{A_j} 
+ \partial_j \vev{\bar{A}_{\bar{i}}} \right) &=& 0, 
\label{eq:abd} \\
\epsilon^{{\rm j}{\rm k}{\rm l}} 
e_{{\rm k}}^{\ k} e_{{\rm l}}^{\ l} 
\partial_k \vev{A_l} &=& 0, 
\label{eq:abfi}
\end{eqnarray}
with $D$ and $F_i$ given by Eqs.~(\ref{eq:osd}) 
and (\ref{eq:osfi}), respectively. 

One of the consequences of nonvanishing magnetic fluxes is 
the YM gauge symmetry breaking. If all the magnetic fluxes 
$M_1^{(i)},M_2^{(i)},\ldots,M_N^{(i)}$ take different values 
from each other, the gauge symmetry is broken down as 
$U(N) \to U(1)^N$. The breaking pattern is changed as 
$U(N) \to \prod_a U(N_a)$ in the case with degenerate magnetic 
fluxes that are written, without loss of generality, as 
\begin{eqnarray}
&M_1^{(i)}=M_2^{(i)}=\cdots=M_{N_1}^{(i)},& 
\nonumber \\
&M_{N_1+1}^{(i)}=M_{N_1+2}^{(i)}=\cdots=M_{N_1+N_2}^{(i)},& 
\nonumber \\
&\vdots& 
\nonumber \\
&M_{N_1+N_2+\cdots+N_{\tilde{N}-1}+1}^{(i)}
=M_{N_1+N_2+\cdots+N_{\tilde{N}-1}+2}^{(i)}=\cdots
=M_{N_1+N_2+\cdots+N_{\tilde{N}-1}+N_{\tilde{N}}}^{(i)},& 
\label{eq:gsb}
\end{eqnarray}
with $\sum_a N_a=N$ and all the fluxes 
$M^{(i)}_{N_1}, M^{(i)}_{N_1+N_2}, \ldots, 
M^{(i)}_{N_1+N_2+ \cdots +N_{\tilde{N}-1}+N_{\tilde{N}}}$ 
take different values from each other. 
The same holds for Wilson-lines, 
$\zeta_1^{(i)},\zeta_2^{(i)},\ldots,\zeta_N^{(i)}$. 
In the following, indices $a,b=1,2,\ldots,\tilde{N}$ 
label the unbroken YM subgroups on the flux and Wilson-line 
background~(\ref{eq:ymvevs}), and traces in expressions 
are performed within such subgroups. 

On the ${\cal N}=1$ supersymmetric toroidal background~(\ref{eq:ymvevs}) 
with the magnetic fluxes~(\ref{eq:abelianflux}) as well as the Wilson-lines 
satisfying Eq.~(\ref{eq:gsb}), 
the zero-modes $(V^{{\bm n}={\bm 0}})_{ab}$ 
of the off-diagonal elements $(V)_{ab}$ ($a \ne b$) 
of the 10D vector superfield $V$ obtain mass terms, 
while the diagonal elements 
$(V^{{\bm n}={\bm 0}})_{aa}$ do not. 
Then, we express the zero-modes 
$(V^{{\bm n}={\bm 0}})_{aa}$, 
which contain 4D gauge fields for the unbroken gauge symmetry 
$\prod_a U(N_a)$, as 
\begin{eqnarray}
(V^{{\bm n}={\bm 0}})_{aa} &\equiv& V^a. 
\nonumber
\end{eqnarray}
On the other hand, for $^\exists j \ne i$ with 
$M^{(j)}_{ab} \equiv M^{(j)}_a - M^{(j)}_b < 0$, 
$M^{(i)}_{ab} \equiv M^{(i)}_a - M^{(i)}_b > 0$ 
and $a \ne b$, 
the zero-mode $(\phi_j^{{\bm n}={\bm 0}})_{ab}$ 
of the off-diagonal element $(\phi_j)_{ab}$ 
of the 10D chiral superfield $\phi_j$ 
degenerates with the number of degeneracy 
$N_{ab}=\prod_k \left| M^{(k)}_{ab} \right|$, 
while $(\phi_j^{{\bm n}={\bm 0}})_{ba}$ 
has no zero-mode solution, 
yielding a 4D supersymmetric chiral generation 
in the $ab$-sector~\cite{Abe:2012ya}. 
The opposite is true for 
$M^{(j)}_{ab} > 0$ and $M^{(i)}_{ab} < 0$ 
yielding a 4D chiral generation in the $ba$-sector. 
Therefore, we denote the zero-mode 
$(\phi_j^{{\bm n}={\bm 0}})_{ab}$ 
with the degeneracy $N_{ab}$ as 
\begin{eqnarray}
(\phi_j^{{\bm n}={\bm 0}})_{ab} &\equiv& 
g\,\phi_j^{{\cal I}_{ab}}, 
\nonumber
\end{eqnarray}
where 
${\cal I}_{ab}$ labels the degeneracy, i.e. generations. 
We normalize $\phi_j^{{\cal I}_{ab}}$ by 
the 10D YM coupling constant $g$.
For more details, see Ref.~\cite{Abe:2012ya} 
and references therein.

\subsection{Three generations induced by magnetic fluxes}
We aim to realize a zero-mode spectrum in 10D SYM theory 
compactified on magnetized tori, that contains the MSSM with 
the gauge symmetry $SU(3)_C \times SU(2)_L \times U(1)_Y$ and 
three generations of the quark and the lepton chiral multiplets, 
by identifying those three with degenerate zero-modes of 
the chiral superfields $\phi_j^{{\cal I}_{ab}}$. 

For such a purpose, we start from the 10D $U(N)$ SYM theory 
with $N=8$ and introduce the following magnetic fluxes 
\begin{eqnarray}
F_{2+2r,3+2r} &=& 2 \pi \left( 
\begin{array}{ccccc}
M^{(r)}_C {\bm 1}_4 & \ & \ \\
\ & M^{(r)}_L {\bm 1}_2 & \ \\
\ & \ & M^{(r)}_R  {\bm 1}_2 
\end{array} \right), 
\end{eqnarray}
where ${\bm 1}_N$ is a $N \times N$ unit matrix, and 
all the nonvanishing entries take different values 
from each other. 
These magnetic fluxes 
break YM symmetry 
as $U(8) \to U(4)_C \times U(2)_L \times U(2)_R$. 
We consider the case that this is further broken down to 
$U(3)_C \times U(1)_{C'} \times U(2)_L 
\times U(1)_{R'} \times U(1)_{R''}$
by the following Wilson lines 
\begin{eqnarray}
\zeta_r &=& \left( 
\begin{array}{ccccc}
\zeta^{(r)}_C {\bm 1}_3 & \ & \ & \ & \ \\
\ & \zeta^{(r)}_{C'} & \ & \ & \ \\
\ & \ & \zeta^{(r)}_L {\bm 1}_2 & \ & \ \\
\ & \ & \ & \zeta^{(r)}_{R'} & \ \\
\ & \ & \ & \ & \zeta^{(r)}_{R''} 
\end{array} \right), 
\label{eq:wlm}
\end{eqnarray}
where 
all the nonvanishing entries take different values 
from each other. 
The gauge symmetries $SU(3)_C$ and $SU(2)_L$ 
of the MSSM are embedded into the above 
unbroken gauge groups as $SU(3)_C \subset U(3)_C$ and 
$SU(2)_L \subset U(2)_L$. 

A combination of the magnetic fluxes, that yield three 
generations from the zero-mode degeneracy and also the 
full-rank Yukawa matrices from the 10D gauge interaction 
as we will see later, are found as 
\begin{eqnarray}
(M^{(1)}_C,M^{(1)}_L,M^{(1)}_R) &=& (0,+3,-3), 
\nonumber \\
(M^{(2)}_C,M^{(2)}_L,M^{(2)}_R) &=& (0,-1,0), 
\nonumber \\
(M^{(3)}_C,M^{(3)}_L,M^{(3)}_R) &=& (0,0,+1), 
\label{eq:model}
\end{eqnarray}
where the supersymmetry conditions 
(\ref{eq:abd}) and (\ref{eq:abfi}) are satisfied by 
\begin{eqnarray}
{\cal A}^{(1)}/{\cal A}^{(2)}
={\cal A}^{(1)}/{\cal A}^{(3)}=3. 
\label{eq:Aratio}
\end{eqnarray}

In this model, chiral superfields 
$Q$, $U$, $D$, $L$, $N$, $E$, $H_u$ and $H_d$ 
carrying 
the left-handed quark, 
the right-handed up-type quark, 
the right-handed down-type quark, 
the left-handed lepton, 
the right-handed neutrino, 
the right-handed electron, 
the up- and the down-type Higgs bosons, respectively, 
are found in $\phi_i^{{\cal I}_{ab}}$ as 
\begin{eqnarray}
\phi_1^{{\cal I}_{ab}} &=& 
\left( 
\begin{array}{cc|c|cc}
\Omega_C^{(1)} & \Xi_{CC'}^{(1)} & 0 & 
\Xi_{CR'}^{(1)} & \Xi_{CR''}^{(1)} \\
\Xi_{C'C}^{(1)} & \Omega_{C'}^{(1)} & 0 & 
\Xi_{C'R'}^{(1)} & \Xi_{C'R''}^{(1)} \\ 
\hline 
\Xi_{LC}^{(1)} & \Xi_{LC'}^{(1)} & \Omega_L^{(1)} & 
H_u^K & H_d^K \\ 
\hline 
0 & 0 & 0 & \Omega_{R'}^{(1)} & \Xi_{R'R''}^{(1)} \\
0 & 0 & 0 & \Xi_{R''R'}^{(1)} & \Omega_{R''}^{(1)} 
\end{array}
\right), 
\nonumber \\
\phi_2^{{\cal I}_{ab}} &=& 
\left( 
\begin{array}{cc|c|cc}
\Omega_C^{(2)} & \Xi_{CC'}^{(2)} & Q^I & 0 & 0 \\
\Xi_{C'C}^{(2)} & \Omega_{C'}^{(2)} & L^I & 0 & 0 \\
\hline 
0 & 0 & \Omega_L^{(2)} & 0 & 0 \\
\hline 
0 & 0 & 0 & \Omega_{R'}^{(2)} & \Xi_{R'R''}^{(2)} \\
0 & 0 & 0 & \Xi_{R''R'}^{(2)} & \Omega_{R''}^{(2)} 
\end{array}
\right), 
\nonumber \\
\phi_3^{{\cal I}_{ab}} &=& 
\left( 
\begin{array}{cc|c|cc}
\Omega_C^{(3)} & \Xi_{CC'}^{(3)} & 0 & 0 & 0 \\
\Xi_{C'C}^{(3)} & \Omega_{C'}^{(3)} & 0 & 0 & 0 \\
\hline 
0 & 0 & \Omega_L^{(3)} & 0 & 0 \\
\hline 
U^J & N^J & 0 & \Omega_{R'}^{(3)} & \Xi_{R'R''}^{(3)} \\
D^J & E^J & 0 & \Xi_{R''R'}^{(3)} & \Omega_{R''}^{(3)} 
\end{array}
\right), 
\label{eq:phicont}
\end{eqnarray}
where the rows and the columns of matrices 
correspond to 
$a=1,\ldots,5=C,C',L,R',R''$ and 
$b=1,\ldots,5=C,C',L,R',R''$, 
respectively, and the indices $I,J=1,2,3$ and 
$K=1,\ldots,6$ label the zero-mode degeneracy, 
i.e., generations. 

Therefore, three generations of 
$Q$, $U$, $D$, $L$, $N$, $E$ and 
six generations of $H_u$ and $H_d$ 
are generated by the magnetic fluxes~(\ref{eq:model}) 
that correspond to 
\begin{eqnarray}
M^{(1)}_C-M^{(1)}_L \ = \ -3, \qquad 
M^{(1)}_L-M^{(1)}_R &=& +6, \qquad 
M^{(1)}_R-M^{(1)}_C \ = \ -3, \nonumber \\
M^{(2)}_C-M^{(2)}_L \ = \ +1, \qquad 
M^{(2)}_L-M^{(2)}_R &=& -1, \qquad 
M^{(2)}_R-M^{(2)}_C \ = \ 0, \nonumber \\
M^{(3)}_C-M^{(3)}_L \ = \ 0, \qquad 
M^{(3)}_L-M^{(3)}_R &=& -1, \qquad 
M^{(3)}_R-M^{(3)}_C \ = \ +1. 
\label{eq:f3gen}
\end{eqnarray}
The zero entries of the matrices in Eq.~(\ref{eq:phicont}) 
represent components eliminated due to the effect of 
chirality projection caused by magnetic fluxes. 
Because some vanishing fluxes are inevitable in Eq.~(\ref{eq:model}) 
in order to realize three generations of quarks and leptons 
with the Yukawa coupling matrices of the full-rank, some of 
$M^{(i)}_{ab}$ become zero in Eq.~(\ref{eq:f3gen}), 
that causes certain massless exotic modes $\Xi_{ab}^{(r)}$ 
as well as massless diagonal components $\Omega_{a}^{(r)}$, 
i.e., the so-called open string moduli, all of which feel zero fluxes. 
These exotics are severely constrained by many experimental 
data at low energies. In the following, we show that most of 
the massless exotic modes can be eliminated if we consider 
a certain orbifold projection on $r=2,3$ tori, that is, 
a sort of magnetized orbifolds~\cite{Abe:2008fi}.

\subsection{Exotic modes and $Z_2$-projection}

Three generations of quarks and leptons are generated in 
the first torus $r=1$ by the magnetic fluxes~(\ref{eq:model}). 
The number of the degenerate zero-modes (generations) 
is changed by the orbifold projection~\cite{Abe:2008fi}.
We assume the $T^6/Z_2$ orbifold where the $Z_2$ acts on the 
second and the third tori $r=2, 3$ in order to eliminate 
only the exotic modes without affecting the generation 
structure of the MSSM matter fields realized by the 
magnetic fluxes~(\ref{eq:model}). 
Then, the $Z_2$ transformation of 10D superfields 
$V$ and $\phi_i$ is assigned for 
$^\forall m=4,5$ and $^\forall n=6,7,8,9$ as 
\begin{eqnarray}
V(x,y_m,-y_n) &=& +P V(x,y_m,-y_n) P^{-1}, 
\nonumber \\
\phi_1(x,y_m,-y_n) &=& +P \phi_1(x,y_m,-y_n) P^{-1}, 
\nonumber \\
\phi_2(x,y_m,-y_n) &=& -P \phi_2(x,y_m,-y_n) P^{-1}, 
\nonumber \\
\phi_3(x,y_m,-y_n) &=& -P \phi_3(x,y_m,-y_n) P^{-1}, 
\label{eq:z2tr}
\end{eqnarray}
where $P$ is a projection operator acting on YM indices 
satisfying $P^2={\bm 1}_N$. 
The $\phi_2$ and $\phi_3$ fields have the minus sign under the 
$Z_2$ reflection, because those are the vector fields, 
$A_i$ ($i=2,3$) on the $Z_2$ orbifold plane.
Note that the orbifold projection~(\ref{eq:z2tr}) respects 
the ${\cal N}=1$ supersymmetry preserved by the magnetic 
fluxes~(\ref{eq:model}), because the $Z_2$-parities are 
assigned to the ${\cal N}=1$ superfields $V$ and $\phi_i$. 

For the matter profile~(\ref{eq:phicont}) caused by 
the magnetic fluxes (\ref{eq:model}), we find that 
the following $Z_2$-projection operator, 
\begin{eqnarray}
P_{ab} &=& \left( 
\begin{array}{ccc}
-{\bm 1}_4 & 0 & 0 \\
0 & +{\bm 1}_2 & 0 \\
0 & 0 & +{\bm 1}_2 
\end{array} \right), 
\nonumber
\end{eqnarray}
removes most of the massless exotic modes $\Xi_{ab}^{(r)}$ 
and some of massless diagonal components $\Omega_{a}^{(r)}$. 
The matter contents on the orbifold $T^6/Z_2$ is found as 
\begin{eqnarray}
\phi_1^{{\cal I}_{ab}} &=& 
\left( 
\begin{array}{cc|c|cc}
\Omega_C^{(1)} & \Xi_{CC'}^{(1)} & 0 & 0 & 0 \\
\Xi_{C'C}^{(1)} & \Omega_{C'}^{(1)} & 0 & 0 & 0 \\ 
\hline 
0 & 0 & \Omega_L^{(1)} & H_u^K & H_d^K \\ 
\hline 
0 & 0 & 0 & \Omega_{R'}^{(1)} & \Xi_{R'R''}^{(1)} \\
0 & 0 & 0 & \Xi_{R''R'}^{(1)} & \Omega_{R''}^{(1)} 
\end{array}
\right), 
\nonumber \\
\phi_2^{{\cal I}_{ab}} &=& 
\left( 
\begin{array}{cc|c|cc}
0 & 0 & Q^I & 0 & 0 \\
0 & 0 & L^I & 0 & 0 \\
\hline 
0 & 0 & 0 & 0 & 0 \\
\hline 
0 & 0 & 0 & 0 & 0 \\
0 & 0 & 0 & 0 & 0 
\end{array}
\right), 
\qquad 
\phi_3^{{\cal I}_{ab}} 
\ = \ 
\left( 
\begin{array}{cc|c|cc}
0 & 0 & 0 & 0 & 0 \\
0 & 0 & 0 & 0 & 0 \\
\hline 
0 & 0 & 0 & 0 & 0 \\
\hline 
U^J & N^J & 0 & 0 & 0 \\
D^J & E^J & 0 & 0 & 0 
\end{array}
\right), 
\nonumber
\end{eqnarray}
where $I,J=1,2,3$ and $K=1,\ldots,6$ 
label the generations as before. 
There still remain massless exotic modes $\Xi_{ab}^{(1)}$ 
for $a,b=C,C'$ and $a,b=R',R''$ with $a \ne b$ as well as 
open string moduli $\Omega_{a}^{(1)}$ for $a=C,C',L,R',R''$. 
That is one of the open problems in the $T^6/Z_2$ magnetized 
orbifold model. In the following phenomenological analyses, 
these exotic modes are assumed to become massive through 
some nonperturbative effects or higher-order corrections, 
so that they decouple from the low-energy physics. 

Due to the orbifold projection~(\ref{eq:z2tr}), 
nonvanishing Wilson-line parameters in Eq.~(\ref{eq:wlm}) 
are possible\footnote{Nonvanishing Wilson-line parameters 
would be possible also in the 2nd and the 3rd tori, 
if we allow non-zero VEVs of vector fields that are 
constants in the bulk but change their sign across the 
fixed points (planes) of the orbifold, that is beyond the 
scope of this paper. In this case localized magnetic fluxes 
at the fixed points might be induced which cause nontrivial 
effects on the wavefunction profile of the charged matter 
fields~\cite{Lee:2003mc}.} 
only in the first torus $r=1$. We denote differences of 
these nonvanishing Wilson-line parameters as 
\begin{eqnarray}
\zeta^{(1)}_C-\zeta^{(1)}_L &\equiv& \zeta_Q, \qquad 
\zeta^{(1)}_{R'}-\zeta^{(1)}_C \ \equiv \ \zeta_U, \qquad 
\zeta^{(1)}_{R''}-\zeta^{(1)}_C \ \equiv \ \zeta_D, 
\nonumber \\
\zeta^{(1)}_{C'}-\zeta^{(1)}_L &\equiv& \zeta_L, \qquad 
\zeta^{(1)}_{R'}-\zeta^{(1)}_{C'} \ \equiv \ \zeta_N, \qquad 
\zeta^{(1)}_{R''}-\zeta^{(1)}_{C'} \ \equiv \ \zeta_E, 
\label{eq:dzeta}
\end{eqnarray}
whose numerical values are determined later phenomenologically.

\subsection{Anomalous $U(1)$s and the hypercharge}

Finally in this section, 
we discuss the $U(1)$ gauge fields and their charges in 
the low energy spectrum. As shown above, most of the exotic 
matter fields become massive (some of them are assumed) 
on the orbifold background, and then the low energy spectrum 
of this model is the MSSM-like matters with additional pairs 
of up and down type Higgs doublets. The gauge group is given 
by $SU(3)_C \times SU(2)_L \times U(1)^5$ and we denote each 
$U(1)_X$ charge as $Q_X$ for $X=a,b,c,d,e$. 
The particle contents and their gauge charges 
are summarized in Table~\ref{tab:fields}. 

\begin{table}[t]
\center
\begin{tabular}{c||c|ccccc|cc}
Matter & $SU(3)_C \times SU(2)_L$ & 
$Q_a$ & $Q_b$ & $Q_c$ & $Q_d$ & $Q_e$ & $Y$ & $B-L$ \\ \hline
$Q$ & $(3,2)$ & $1$ & 0 & $-1$ & 0 & 0 & $1/6$ & 1/3 \\
$U$ & $(\bar{3},2)$ & $-1$ & $0$ & 0 & 1 & 0 & $-2/3$ & $-1/3$ \\
$D$ & $(\bar{3},2)$ & $-1$ & $0$ & 0 & 0 & 1 & $1/3$ & $-1/3$ \\
$L$ & $(1,2)$ & $0$ & 1 & $-1$ & 0 & 0 & $-1/2$ & $-1$ \\
$N$ & $(1,1)$ & $0$ & $-1$ & 0 & 1 & 0 & $0$ & 1 \\
$E$ & $(1,1)$ & $0$ & $-1$ & 0 & 0 & 1 & $1$ & 1 \\
$H_u$ & $(1,2)$ & 0 & 0 & 1 & $-1$ &  0 & $+1/2$ & 0 \\
$H_d$ & $(1,2)$ & 0 & 0 & 1 &  0 & $-1$ & $-1/2$ & 0 \\
\end{tabular}
\caption{Matter fields and their gauge charges.}
\label{tab:fields}
\end{table}

As is well known, there are two non-anomalous local and global 
U(1) symmetries in the MSSM that we denote $U(1)_Y$ and $U(1)_{B-L}$, 
respectively, where $Y$ represents the hypercharge and $B$ ($L$) 
is the baryon (lepton) charge. In our model the $U(1)_{B-L}$ is 
a local symmetry, and there is an additional anomaly-free U(1) 
symmetry as a linear combination of all the U(1) groups denoted by 
$U(1)_D$ with the charge defined by $Q_D=Q_a+Q_b+Q_c+Q_d+Q_e$. 
Each $U(1)_X$ gauge symmetry has clear interpretation in terms 
of the global symmetries in the MSSM. For example, $Q_a$ is related 
to the baryon number and $Q_c$ is nothing but the lepton number. 
Thus one can obtain the $U(1)_{B-L}$ and the $U(1)_Y$ as linear 
combinations of the above five U(1) gauge symmetries. 
Here we take the $U(1)_{B-L}$ charge $Q_{B-L}$ as 
\begin{eqnarray}
Q_{B-L} &=& \frac{1}{3}Q_a - Q_b.
\nonumber
\end{eqnarray}
The U(1) hypercharge $Q_{Y}$ can be given by 
\begin{eqnarray}
Q_{Y} &=& \alpha Q_a 
+ \left( \alpha - \frac{2}{3} \right) Q_b 
+ \left( \alpha - \frac{1}{6} \right) Q_c 
+ \left( \alpha - \frac{2}{3} \right) Q_d 
+ \left( \alpha + \frac{1}{3} \right) Q_e, 
\nonumber
\end{eqnarray}
where $\alpha$ is an arbitrary number. 
It is easy to check that these three U(1) symmetries, 
$U(1)_Y$, $U(1)_{B-L}$ and $U(1)_D$, are anomaly-free 
for both the mixed U(1) and non-Abelian gauge groups. 
As for $U(1)_D$ gauge group, there is no charged chiral matter 
under this gauge group, so the $U(1)_D$ gauge field can decouple. 

In the following, we assume that the $U(1)_{B-L}$ 
is spontaneously broken at a high energy scale. 
There is another $U(1)$ gauge symmetry which 
has a property of Peccei-Quinn symmetry $U(1)_{\rm PQ}$, 
whose charges for matter and Higgs fields are $-1/2$ and $+1$, 
respectively. 
This $U(1)_{\rm PQ}$ symmetry prohibit the so-called $\mu$-term. 
However the $U(1)_{\rm PQ}$ symmetry as well as the remaining 
fifth $U(1)$ symmetry are anomalous, and then we assume 
all the gauge fields of the anomalous $U(1)$s become massive via, 
e.g., the Green-Schwarz mechanism~\cite{Green:1984sg}, 
and decouple from the low-energy physics. 
Then, it is interesting to survey the possibility of the 
dynamical generation of the $\mu$-term, although we just 
assume its existence in the following phenomenological analysis.

\section{Flavor structures of (s)quarks and (s)leptons}
\label{sec:flavor}

In this section we show that a semi-realistic pattern of 
quark and lepton mass matrices are realized at a certain 
point of the (tree-level) moduli space in our model. 
The hierarchical structure of the Yukawa 
couplings is achieved by the wavefunction localization of 
the matter fields in extra dimensions, whose localization 
profiles are completely determined by the magnetic 
fluxes~(\ref{eq:f3gen}). More interestingly, if we embed 
the 10D SYM theory, which is the starting point of our model, into 
10D supergravity, the flavor structure of the superparticles 
induced by the moduli-mediated supersymmetry breaking is also 
fully determined by the wavefunction profile. Therefore the 
supersymmetric flavor structure of the model can be analyzed 
based on the effective supergravity action derived through a 
systematic way proposed in Ref.~\cite{Abe:2012ya}. 

The 4D effective action with the 
${\cal N}=1$ local supersymmetry is generally written in terms 
of 4D ${\cal N}=1$ conformal supergravity~\cite{Kaku:1977rk} as 
\begin{eqnarray}
S_{{\cal N}=1} &=& \int d^4x \sqrt{-g^C} \Bigg[ 
-3 \int d^4 \theta\,\bar{C}C \, e^{-K/3} 
\nonumber \\ && \qquad \qquad \qquad 
+\left\{ \int d^2 \theta \left( 
\frac{1}{4} \sum_a f_a\, W^{a,\alpha} W^a_\alpha 
+C^3 W\right) +{\rm h.c.} \right\} \Bigg], 
\label{eq:csugra}
\end{eqnarray}
where $K$, $W$ and $f_a$ are the effective K\"ahler potential, 
the superpotential and the gauge kinetic functions, respectively, 
as functions of light modes as well as the moduli, and the chiral 
superfield $C$ plays a role of superconformal compensator. 
Here and hereafter, we work in a unit that the 4D Planck scale is unity.

\subsection{The MSSM sector in the 4D effective theory}

The effective 
K\"ahler potential $K$, 
superpotential $W$ and 
gauge kinetic functions $f_a$ ($a=1,2,3$) 
for the MSSM sector of our model on the $T^6/Z_2$ 
magnetized orbifold at the leading order are found in 
the 4D effective action~(\ref{eq:csugra}) as~\cite{Abe:2012ya} 
\begin{eqnarray}
K &=& K^{(0)}(\bar\Phi^{\bar{m}},\Phi^m) 
+Z^{({\cal Q})}_{\bar{\cal I}{\cal J}}(\bar\Phi^{\bar{m}},\Phi^m) 
\bar{\cal Q}^{\bar{\cal I}} {\cal Q}^{\cal J}, 
\nonumber \\ 
W &=& \lambda^{({\cal Q})}_{{\cal I}{\cal J}{\cal K}}(\Phi^m) 
{\cal Q}^{\cal I} {\cal Q}^{\cal J} {\cal Q}^{\cal K}, 
\nonumber \\ 
f_a &=& S \quad (a= 1, 2, 3), 
\label{eq:mssm}
\end{eqnarray}
where ${\cal Q}^{\cal I}$ and $\Phi^m$ symbolically represents 
the MSSM matter and the moduli chiral superfields, 
\begin{eqnarray}
{\cal Q}^{\cal I} &=& 
\{ Q^I, U^J, D^J, L^I, N^J, E^J, H_u^K, H_d^K \}, \qquad 
\Phi^m \ = \ \{ S, T_r, U_r \}, 
\label{eq:matter}
\end{eqnarray}
respectively, the subscript $r=1,2,3$ labels the $r$th 
two-dimensional torus $T^2$ among the factorizable 
three tori $T^2 \times T^2 \times T^2$, 
and traces of the YM-indices are implicit. 
The explicit expressions of 
the moduli K\"ahler potential 
$K^{(0)}(\bar\Phi^{\bar{m}},\Phi^m)$, 
the matter K\"ahler metrics 
$Z^{({\cal Q})}_{\bar{\cal I}{\cal J}}(\bar\Phi^{\bar{m}},\Phi^m)$ 
and the holomorphic Yukawa couplings 
$\lambda^{({\cal Q})}_{{\cal I}{\cal J}{\cal K}}(\Phi^m)$ 
are exhibited in Appendix~\ref{app:kmhy}. 

We assume a certain mechanism of moduli 
stabilization and supersymmetry breaking 
that fixes VEVs of moduli superfields, 
\begin{eqnarray}
\langle S \rangle &\equiv& 
s + \theta^2 F^S, \qquad 
\langle T_r \rangle \ \equiv \ 
t_r + \theta^2 F^T_r, \qquad 
\langle U_r \rangle \ \equiv \ 
u_r + \theta^2 F^U_r. 
\nonumber
\end{eqnarray}
Note that these VEVs determine 
10D parameters $g$, ${\cal A}^{(i)}$ and $\tau_i$ 
as~\cite{Cremades:2004wa} 
\begin{eqnarray}
{\rm Re}\,s &=& g^{-2} \prod_{r=1}^3 {\cal A}^{(r)}, 
\qquad 
{\rm Re}\,t_r \ = \ g^{-2} {\cal A}^{(r)}, 
\qquad 
u_r \ = \ i \bar\tau_r, 
\label{eq:stuvev}
\end{eqnarray}
and then $S$, $T_r$ and $U_r$ are called 
the dilaton, the K\"ahler moduli and 
the complex structure moduli, respectively, 
in the 4D effective theory. 
In the following analyses, numerical values of 
the moduli VEVs as well as 
 the Wilson-lines are scanned 
phenomenologically.

\subsection{Quark and lepton masses and mixings}

First we analyze the flavor structure of the SM sector 
in our model. Canonically normalized Yukawa couplings 
between three generations of the quarks or the leptons 
and six generations of the Higgs doublets are calculated by 
\begin{eqnarray}
y^{(U)}_{IJK} &=& 
\frac{\lambda^{(U)}_{IJK}}{
\sqrt{Y^{({Q})}_{\bar{I}I} 
Y^{({U})}_{\bar{J}J} 
Y^{({H}_u)}_{\bar{K}K}}}, \qquad 
 y^{({D})}_{IJK} = 
\frac{\lambda^{({D)}_{IJK}}}{
\sqrt{Y^{({Q})}_{\bar{I}I} 
Y^{({D})}_{\bar{J}J} 
Y^{({H}_d)}_{\bar{K}K}}},   \nonumber \\
y^{(N)}_{IJK} &=& 
\frac{\lambda^{(N)}_{IJK}}{
\sqrt{Y^{({L})}_{\bar{I}I} 
Y^{({N})}_{\bar{J}J} 
Y^{({H}_u)}_{\bar{K}K}}}, \qquad 
 y^{({E})}_{IJK} = 
\frac{\lambda^{({E)}_{IJK}}}{
\sqrt{Y^{({L})}_{\bar{I}I} 
Y^{({E})}_{\bar{J}J} 
Y^{({H}_d)}_{\bar{K}K}}},
\label{eq:yukawa}
\end{eqnarray}
where 
$Y^{({\cal Q})}_{\bar{\cal I}{\cal J}}$ represents 
the superspace wavefunction coefficient of 
$\bar{\cal Q}^{\bar{\cal I}} {\cal Q}^{{\cal J}}$ 
in the superspace action, 
which is related to the K\"ahler metric as 
\begin{eqnarray}
Y^{({\cal Q})}_{\bar{\cal I}{\cal J}} &=& 
e^{-K_0(\bar\Phi^{\bar{m}},\Phi^m)/3} 
Z^{({\cal Q})}_{\bar{\cal I}{\cal J}}
(\bar\Phi^{\bar{m}},\Phi^m). 
\nonumber
\end{eqnarray}
The above Yukawa coupling~(\ref{eq:yukawa}) possesses 
the flavor symmetry $\Delta(27)$~\cite{Abe:2009vi} 
due to the choice of the magnetic fluxes~(\ref{eq:model}) 
selected for the three generations of quarks and leptons 
with the full-rank Yukawa matrices. 

The up- and down-type quark masses 
$(m_u, m_c, m_t)$ and $(m_d, m_s, m_b)$, 
the neutrino Dirac masses 
$(m_{\nu_e}, m_{\nu_\mu}, m_{\nu_\tau})$ and 
the charged lepton masses $(m_e, m_\mu, m_\tau)$ are 
the eigenvalues of the $3 \times 3$ mass matrices 
\begin{eqnarray}
y^{(U)}_{IJK} \langle H_u^K \rangle 
&\equiv& y^{(U)}_{IJ} v_u, \qquad 
y^{(D)}_{IJK} \langle H_d^K \rangle 
\ \equiv \ y^{(D)}_{IJ} v_d, \nonumber \\
y^{(N)}_{IJK} \langle H_u^K \rangle 
&\equiv& y^{(N)}_{IJ} v_u, \qquad 
y^{(E)}_{IJK} \langle H_d^K \rangle 
\ \equiv \ y^{(E)}_{IJ} v_d, 
\nonumber
\end{eqnarray}
respectively, that is, 
\begin{eqnarray}
({V^{(U)}_L}^\dagger)_{\hat{I}}^{\ I}\,
y^{(U)}_{IJ} \, (V^{(U)}_R)^J_{\ \hat{J}} 
&=& {\rm diag}\,(m_u, m_c, m_t)_{\hat{I}\hat{J}}/v_u, 
\nonumber \\
({V^{(D)}_L}^\dagger)_{\hat{I}}^{\ I}\, 
y^{(D)}_{IJ} \, (V^{(D)}_R)^J_{\ \hat{J}} 
&=& {\rm diag}\,(m_d, m_s, m_b)_{\hat{I}\hat{J}}/v_d, 
\nonumber \\
({V^{(N)}_L}^\dagger)_{\hat{I}}^{\ I}\,
y^{(N)}_{IJ} \, (V^{(N)}_R)^J_{\ \hat{J}} 
&=& {\rm diag}\,(m_{\nu_1}, m_{\nu_2}, 
m_{\nu_3})_{\hat{I}\hat{J}}/v_u, 
\nonumber \\
({V^{(E)}_L}^\dagger)_{\hat{I}}^{\ I}\,
y^{(E)}_{IJ} \, (V^{(E)}_R)^J_{\ \hat{J}} 
&=& {\rm diag}\,(m_e, m_\mu, m_\tau)_{\hat{I}\hat{J}}/v_d, 
\label{eq:mevs}
\end{eqnarray}
where $V^{({\cal Q}_y)}_{L,R}$ for ${\cal Q}_y = U, D, N, E$ 
are unitary matrices, and $\hat{I}, \hat{J} = 1, 2, 3$ 
label the (Dirac) mass eigenstates. 
The Cabibbo-Kobayashi-Maskawa (CKM) matrix~\cite{Kobayashi:1973fv} 
$V_{\rm CKM} \equiv V^{(U)}_L {V^{(D)}_L}^\dagger$ 
describes the flavor mixing in the quark sector, 
whose matrix elements are precisely measured by experiments. 

For $v_u=v \sin \beta$, $v_d= v \cos \beta$ and $v=174$ GeV, 
we numerically find that the following values of the $\tan\beta$ 
and the VEVs of Higgs fields, 
\begin{eqnarray}
\tan\beta &=& 25 \\
\vev{H_u^K} &=& (0.0,\,0.0,\,2.7,\,1.3,\,0.0,\,0.0)\,v_u\,
\times\,{\cal N}_{H_u}, 
\nonumber \\
\vev{H_d^K} &=& (0.0,\,0.1,\,5.8,\,5.8,\,0.0,\,0.1)\,v_d\,
\times\,{\cal N}_{H_d}, 
\label{eq:higgsvevs}
\end{eqnarray}
and those of the geometric moduli~(\ref{eq:stuvev}) 
as well as the Wilson-line parameters~(\ref{eq:dzeta}), 
\begin{eqnarray}
\pi s &=& 6.0, \nonumber \\
(t_1,t_2,t_3) &=& (3.0,\,1.0,\,1.0)\,\times\,2.8\,\times\,10^{-8}, 
\nonumber \\
(\tau_1,\tau_2,\tau_3) &=& (4.1i,\,1.0i,\,1.0i), 
\nonumber \\
(\zeta_Q,\zeta_U,\zeta_D,\zeta_L,\zeta_N,\zeta_E) 
&=& (1.0i,\,1.9i,\,1.4i,\,0.7i,\,2.2i,\,1.7i), 
\label{eq:modelwl}
\end{eqnarray}
yield a semi-realistic pattern of the quark and the charged lepton 
masses as well as the CKM matrix at the electroweak (EW) scale shown 
in Table~\ref{tab:ckm}. The normalization factors 
${\cal N}_{H_u}=1/\sqrt{2.7^2+1.3^2}$ and 
${\cal N}_{H_d}=1/\sqrt{2(0.1^2+5.8^2)}$ in Eq.~(\ref{eq:higgsvevs}) 
are factorized just for convenience later in Eq.~(\ref{eq:mukl}).

Here, we assume some nonperturbative 
effects~\cite{Ibanez:2006da} and/or higher-dimensional operators 
that effectively generate supersymmetric mass terms, 
\begin{eqnarray}
W_{\rm eff} &=& \mu_{KL} H_u^K H_d^L .
\label{eq:mu}
\end{eqnarray}
Because VEVs of these Higgs fields shown in Eq.~(\ref{eq:higgsvevs}) 
generate a semi-realistic pattern of the quark and the lepton 
masses and their mixing angles, here we consider the case that the supersymmetric 
mass parameters $\mu_{KL}$ are aligned in such a way that 
\begin{eqnarray}
\sum_{K,L}\,(U_{H_u})_{\hat{K}}^{\ K}\, 
\mu_{KL}\,(U_{H_d}^\dagger)^{L}_{\ \hat{L}}
&=& \delta_{\hat{K} \hat{L}}\,\mu_{\hat{K}}, \qquad 
|\mu_{\hat{K}=1}| \ll M_{\rm GUT} \lesssim |\mu_{\hat{K} \ne 1}|, 
\nonumber \\
(U_{H_{u,d}})_{\hat{K}=1}^{\ K} 
&=& \langle H_{u,d}^K \rangle/v_{u,d}, 
\label{eq:mukl}
\end{eqnarray}
are satisfied with unitary matrices $U_{H_{u,d}}$, 
where 
$\hat{K}, \hat{L} = 1, 2, \ldots, 6$ label the supersymmetric 
mass eigenstates diagonalizing $\mu_{KL}$, 
and the VEVs $\langle H_{u,d}^K \rangle$ represent 
those shown in Eq.~(\ref{eq:higgsvevs}). In this case, 
five of the six Higgs doublets other than $H_{u,d}^{\hat{K}=1}$ 
decouple from the light modes due to the heavy supersymmetric 
masses $\mu_{\hat{K} \ne 1}$. 
In the following, the numerical value of the $\mu$-parameter 
$\mu \equiv \mu_{\hat{K}=1}$
is determined so that the EW symmetry is broken successfully 
yielding the observed masses of the $W$ and the $Z$ bosons, 
and then the masses and the mixing angles of quarks and leptons 
shown in Tables~\ref{tab:ckm}  and \ref{tab:mns} are realized.

The VEV of dilaton $\pi s =6.0$ yields the unified gauge couplings 
$4\pi /{g_a}^2 =24$ at the GUT scale 
$M_{\rm GUT} = 2.0 \times 10^{16}$ GeV, that is implemented 
in the MSSM with low energy data. 
We select the overall magnitudes of $t_r$ so that the compactification 
scale, i.e., the mass scale of the lightest Kaluza-Klein mode 
becomes as high 
as $M_{\rm GUT}$, and their ratios are 
defined to preserve supersymmetry conditions~(\ref{eq:Aratio}). 
Here, the running of the parameters from $M_{\rm GUT}$ 
to the EW scale is evaluated by the one-loop renormalization 
group (RG) equations of the MSSM. 
From Table~\ref{tab:ckm}, we find that the observed hierarchies 
among three generations of quarks and charged leptons are 
realized even with the above non-hierarchical VEVs of 
fields~(\ref{eq:higgsvevs}) and (\ref{eq:modelwl}). 
It is quite interesting and suggestive that the complicated 
flavor structure of our real world could be realized at a 
certain point in the (tree-level) moduli space of the 10D 
SYM theory, whose action is simply given by Eq.~(\ref{eq:10dsym}) 
at the leading order in a rigid limit. 

\begin{table}[t]
\begin{center}
\begin{tabular}{|c||c|c|} \hline
 & Sample values & Observed \\ \hline
$(m_u, m_c, m_t)$ & 
$(3.1 \times 10^{-3}, 1.01, 1.70 \times 10^2)$ & 
$(2.3 \times 10^{-3}, 1.28, 1.74 \times 10^2)$ 
\\ \hline
$(m_d, m_s, m_b)$ & 
$(2.8 \times 10^{-3}, 1.48 \times 10^{-1}, 6.46)$ & 
$(4.8 \times 10^{-3}, 0.95 \times 10^{-1}, 4.18)$ 
\\ \hline
$(m_e, m_\mu, m_\tau)$ & 
$(4.68 \times 10^{-4}, 5.76 \times 10^{-2}, 3.31)$ & 
$(5.11 \times 10^{-4}, 1.06 \times 10^{-1}, 1.78)$ 
\\ \hline \hline 
$|V_{\rm CKM}|$ & 
\begin{minipage}{0.3\linewidth}
\begin{eqnarray} 
\left( 
\begin{array}{ccc}
0.98 & 0.21 & 0.0023 \\
0.21 & 0.98 & 0.041 \\
0.011 & 0.040 & 1.0 
\end{array}
\right) 
\nonumber
\end{eqnarray} \\*[-20pt]
\end{minipage}
& 
\begin{minipage}{0.3\linewidth}
\begin{eqnarray} 
\left( 
\begin{array}{ccc}
0.97 & 0.23 & 0.0035 \\
0.23 & 0.97 & 0.041 \\
0.0087 & 0.040 & 1.0 
\end{array}
\right) 
\nonumber
\end{eqnarray} \\*[-20pt]
\end{minipage} \\ \hline
\end{tabular}
\end{center}
\caption{
Numerical values of the quark masses 
($m_u$, $m_c$, $m_t$), ($m_d$, $m_s$, $m_b$)
and the charged lepton masses 
($m_e$, $m_\mu$, $m_\tau$) 
as well as the absolute values of the elements in 
the CKM matrix $V_{\rm CKM}$ at the EW scale, 
evaluated at a sample point in the moduli space 
of the 10D SYM theory identified by 
the magnetic fluxes~(\ref{eq:model}), 
the Wilson-lines 
and the VEVs of the moduli~(\ref{eq:modelwl}) 
and the Higgs fields~(\ref{eq:higgsvevs}). 
The experimental data~\cite{Beringer:1900zz} are also shown. 
All the mass scales are measured in the unit of GeV.} 
\label{tab:ckm}
\end{table}

In addition, if we assume some nonperturbative 
effects~\cite{Ibanez:2006da} 
or higher-dimensional operators effectively generate 
Majorana masses\footnote{
Note that the Majorana mass term violates the flavor symmetry 
of the Yukawa couplings mentioned above. This fact can be a 
guiding principle for identifying an origin of the term.} 
for the right-handed neutrino $N^J$ in the superpotential such as 
\begin{eqnarray}
W_{\rm eff} &=& M^{(N)}_{IJ} N^I N^J, 
\label{eq:wmm}
\end{eqnarray}
a numerical value of the Majorana mass matrix $M^{(N)}$ is found as 
\begin{eqnarray}
M^{(N)} &=& \left( \begin{array}{ccc} 
1.1 & 1.3 & 0 \\ 1.3 & 0 & 3.2 \\ 0 & 3.2 & 1.8 
\end{array} \right) \ \times \ 10^{12} 
\textrm{\ GeV}, 
\label{eq:majorana}
\end{eqnarray}
that yields a semi-realistic pattern of 
the neutrino masses and 
the Pontecorvo-Maki-Nakagawa-Sakata (PMNS) 
lepton mixing matrix~\cite{Pontecorvo:1967fh} 
at the EW scale shown in Table~\ref{tab:mns}. 

\begin{table}[t]
\begin{center}
\begin{tabular}{|c||c|c|} \hline
 & Sample values & Observed 
\\ \hline
$(m_{\nu_1}, m_{\nu_2}, m_{\nu_3})$ & 
$(3.6 \times 10^{-19}, 8.8 \times 10^{-12}, 2.7 \times 10^{-11})$ & 
$< \ 2 \times 10^{-9}$ 
\\ \hline
$|m_{\nu_1}^2-m_{\nu_2}^2|$ & 
$7.67 \times 10^{-23}$ & 
$7.50 \times 10^{-23}$ 
\\ \hline
$|m_{\nu_1}^2-m_{\nu_3}^2|$ & 
$7.12 \times 10^{-22}$ & 
$2.32 \times 10^{-21}$ 
\\ \hline \hline 
$|V_{\rm PMNS}|$ & 
\begin{minipage}{0.3\linewidth}
\begin{eqnarray} 
\left( 
\begin{array}{ccc}
0.85 & 0.46 & 0.25 \\
0.50 & 0.59 & 0.63 \\
0.15 & 0.66 & 0.73 
\end{array}
\right) 
\nonumber
\end{eqnarray} \\*[-20pt]
\end{minipage}
& 
\begin{minipage}{0.3\linewidth}
\begin{eqnarray} 
\left( 
\begin{array}{ccc}
0.82 & 0.55 & 0.16 \\
0.51 & 0.58 & 0.64 \\
0.26 & 0.61 & 0.75 
\end{array}
\right) 
\nonumber
\end{eqnarray} \\*[-20pt]
\end{minipage} \\ \hline
\end{tabular}
\end{center}
\caption{
Numerical values of the neutrino masses 
($m_{\nu_1}$, $m_{\nu_2}$, $m_{\nu_3}$) 
as well as the absolute values of the 
elements in the PMNS matrix $V_{\rm PMNS}$ at the EW scale, 
evaluated at the same sample point in the moduli space 
as in Table~\ref{tab:ckm} 
but with the Majorana masses~(\ref{eq:majorana}). 
The experimental data~\cite{Beringer:1900zz} are also shown. 
All the mass scales are measured in the unit of GeV.} 
\label{tab:mns}
\end{table}

\subsection{Soft supersymmetry breaking terms}

The low energy features of the superparticles in our model are 
governed by soft supersymmetry breaking parameters, namely, 
the gaugino masses $M_a$, 
the scalar masses $(m_{{\cal Q}}^2)_{\bar{\cal I}{\cal J}}$ 
and the scalar trilinear couplings $A^{({\cal Q}_y)}_{IJK}$ 
(normalized by the corresponding Yukawa couplings), 
those appear in the soft supersymmetry breaking terms as 
\begin{eqnarray}
{\cal L}_{\rm soft} &=& 
-\frac{1}{2} M_a \lambda^a \lambda^a 
-(m^2_{\tilde{\cal Q}})_{\bar{\cal I}{\cal J}} 
(\tilde{\cal Q}^{\cal I})^\dagger \tilde{\cal Q}^{\cal J} 
-\frac{1}{6} \sum_{\{ {\cal Q}_y \}}
y^{({\cal Q}_y)}_{IJK} 
A^{({\cal Q}_y)}_{IJK} 
\tilde{\cal Q}^I_L 
\tilde{\cal Q}^J_R 
\tilde{\cal Q}^K_H 
+{\rm h.c.}, 
\nonumber
\end{eqnarray}
where the superscript ${\cal Q}_y$ represents 
${\cal Q}_y = U, D, N, E$ and 
$({\cal Q}_L,{\cal Q}_R, {\cal Q}_H) = (Q,U,H_u)$, 
$(Q,D,H_d)$, $(L,N,H_u)$ and $(L,E,H_d)$ for 
${\cal Q}_y = U, D, N$ and $E$, respectively.
The tilded fields $\tilde{\cal Q}^{\cal I}$ denote 
the scalar fields, 
\begin{eqnarray}
\tilde{\cal Q}^{\cal I} &=& 
\{ \tilde{q}^I, \tilde{u}^J, \tilde{d}^J, 
\tilde{l}^I, \tilde{\nu}^J, \tilde{e}^J, 
h_u^K, h_d^K \}, 
\nonumber
\end{eqnarray}
those are the lowest components of the chiral superfields 
${\cal Q}^{\cal I}$ in the $\theta$ and the $\bar\theta$ 
expansion. 
Note that only the direction with $\hat K =1$ remains light 
in the Higgs sector of $H^K_u$ and $H^K_d$. 
The so-called $B$-term also appears as the soft supersymmetry breaking
term.
In the following, its value is determined numerically such that 
the EW symmetry is broken successfully.

The explicit moduli dependence of the K\"ahler and 
the superpotential~(\ref{eq:mssm}) in the MSSM sector 
allows us to determine moduli-mediated contributions
\cite{Kaplunovsky:1993rd}  to 
the soft supersymmetry breaking parameters (induced by 
nonvanishing $F$-components of $S$, $T_r$ and $U_r$) 
as well as the anomaly-mediated one \cite{Randall:1998uk} 
(induced by a nonvanishing $F$-component of $C$). 
These contributions are summarized as~\cite{Choi:2005ge} 
\begin{eqnarray}
M_a &=& F^m \partial_m \ln ({\rm Re} f_a) 
+\frac{b_a g_a^2}{8 \pi^2} \frac{F^C}{C_0}, 
\nonumber \\
(m_{\tilde{\cal Q}}^2)_{\bar{\cal I}{\cal J}} &=& 
-F^m \bar{F}^{\bar{n}} 
\partial_m \partial_{\bar{n}} 
\ln Y^{({\cal Q})}_{\bar{\cal I}{\cal J}}
-\frac{\delta_{\bar{\cal I}{\cal J}}}{32 \pi^2} 
\frac{d \gamma_{{\cal Q}^{\cal J}}}{d \ln \mu} 
\left| \frac{F^C}{C_0} \right|^2 
+\frac{\delta_{\bar{\cal I}{\cal J}}}{16 \pi^2} \left( 
\frac{\bar{F}^{\bar{C}}}{\bar{C}_0} 
F^m \partial_m \gamma_{{\cal Q}^{\cal J}} 
+{\rm h.c.} \right), 
\nonumber \\
A^{({\cal Q}_y)}_{IJK} &=& 
-F^m \partial_m \ln \left( 
\frac{\lambda^{({\cal Q}_y)}_{IJK}}{
Y^{({\cal Q}_L)}_{\bar{I}I} 
Y^{({\cal Q}_R)}_{\bar{J}J} 
Y^{({\cal Q}_H)}_{\bar{K}K}} \right) 
-\frac{
\gamma_{{\cal Q}_L^I} +
\gamma_{{\cal Q}_R^J} +
\gamma_{{\cal Q}_H^K}}{16 \pi^2} 
\frac{F^C}{C_0}, 
\label{eq:softp}
\end{eqnarray}
where $\gamma_{{\cal Q}^{\cal J}}$ is the anomalous 
dimension of ${\cal Q}^{{\cal J}}$, and 
$F^m$ represents $F$-components 
of moduli superfields, that is, 
\begin{eqnarray}
F^m &=& \{ F^S, F^T_r, F^U_r \}, 
\nonumber
\end{eqnarray}
while $C_0$ and $F^C$ are the lowest 
and the $\theta^2$ components of $C$, respectively, 
in the $\theta$ and the $\bar\theta$ expansion. Here, 
we fix the dilatation symmetry by $C_0 = \exp(K|_{\theta=\bar\theta=0}/6)$ 
that corresponds to the Einstein frame. 

In the following, we study phenomenological aspects of 
our model at low energies, in the case that the above 
soft parameters are dominated by the moduli- and the 
anomaly-mediated contributions and the other contributions 
(such as the gauge-mediated one that is further model dependent) 
are negligible, by assuming a certain moduli stabilization and a 
supersymmetry breaking mechanism outside the MSSM 
sector that cause such a situation.

\section{Phenomenological aspects at low energies}
\label{sec:pheno}

It has been found that the three generations of quarks and leptons 
are obtained from the degeneracy of chiral zero-modes due to the 
magnetic fluxes~(\ref{eq:model}), yielding consequently the 
six-generations of up- and down-type Higgs doublets. 
Furthermore, a semi-realistic pattern 
of the quark and the charged lepton masses and the CKM mixings 
can be realized  as shown in Tables~\ref{tab:ckm} 
at a certain point in the moduli space of the 10D SYM theory 
where the numerical values of the Higgs and the moduli VEVs 
as well as the Wilson-line parameters are given as shown in 
Eqs.~(\ref{eq:higgsvevs}) and (\ref{eq:modelwl}).

The undetermined parameters so far are supersymmetry 
breaking order parameters $F^m=\{ F^S, F^T_r, F^U_r\}$ and 
$F^C$ mediated by moduli and compensator chiral multiplets 
$\Phi^m=\{S, T_r, U_r \}$ and $C$, respectively. As a 
representative scale of the supersymmetry breaking $M_{\rm SB}$, 
we refer the $F$-component of the dilaton superfield $S$, 
\begin{eqnarray}
M_{\rm SB} &\equiv& \sqrt{K_{S \bar{S}}}\,F^S, 
\nonumber
\end{eqnarray}
and define ratios, 
\begin{eqnarray}
R^T_r &=& 
\frac{\sqrt{K_{T_r \bar{T}_r}}\,F^T_r}{M_{\rm SB}}, 
\qquad 
R^U_r \ = \ 
\frac{\sqrt{K_{U_r \bar{U}_r}}\,F^U_r}{M_{\rm SB}}, 
\qquad 
R^C \ = \ \frac{1}{4 \pi^2} 
\frac{F^C/C_0}{M_{\rm SB}}. 
\label{eq:rtuc}
\end{eqnarray}
Here, we assume that CP phases of $F^S$, $F^T_r$, $F^U_r$ 
and $F^C$ are the same, and $R^T_r$, $R^U_r$ and $R^C$ 
are real.
Then, there is no physical CP violation due to supersymmetry breaking 
terms.
Otherwise, there would be a strong constraint on 
CP violation in the soft supersymmetry breaking terms.

As shown in Eq.~(\ref{eq:mssm}), 
the gauge kinetic functions depend on only the dilaton 
superfield $S$ at the tree level in the (leading order) 
effective supergravity action\footnote{
A moduli mixing in the gauge kinetic functions could 
occur due to higher-order corrections if the SYM theory 
originates, e.g., from (magnetized) D-branes~\cite{Lust:2004cx} 
that is beyond the scope of this paper.}. 
Then, the gaugino masses shown in Eq.~(\ref{eq:softp}) 
are determined by $F^S$ and $F^C$ at the compactification 
scale independently of $R^T_r$ and $R^U_r$. 
On the other hand, the lower bound on the gluino mass 
$M_3 \gtrsim 860$ GeV is found from the recent LHC data~\cite{:2012rz}. 
In the following we analyze phenomenological features of 
our model for $M_{\rm SB} = 1$ TeV satisfying the above condition.
By varying the ratios $R^T_r$, $R^U_r$ and $R^C$, 
we show phenomenological aspects of our model, especially, 
typical sizes of the flavor violations caused by superparticles. 

Hereafter, we neglect all the Yukawa couplings except for 
those involving only the third generations, 
$y^{(U)}_{33}$, $y^{(D)}_{33}$, $y^{(N)}_{33}$ and $y^{(E)}_{33}$, 
for a numerical performance, 
when we evaluate soft parameters and their RG running. 
We also reevaluate accordingly the RG runnings of Yukawa couplings 
in this approximation that was not adopted in the analysis of quark 
and lepton masses and mixings.

\subsection{Supersymmetric flavor violations}

In models with a low-energy supersymmetry breaking, the flavor 
violations such as FCNCs caused by superparticles are severely 
constrained by the experiments. As measures of such supersymmetric 
flavor violations in our model, 
we adopt so-called mass insertion parameters~\cite{Misiak:1997ei}, 
those we define as 
\begin{eqnarray}
(\delta^{({\cal Q}_y)}_{LR})_{IJ} 
&\equiv& \frac{v_f \left( 
V^{({\cal Q}_y)}_L 
{a^{({\cal Q}_y)}}^\dagger 
{V^{({\cal Q}_y)}_R}^\dagger \right)_{IJ}}{
\sqrt{ 
\left( V^{({\cal Q}_y)}_L 
m^2_{\tilde{\cal Q}_L} 
{V^{({\cal Q}_y)}_L}^\dagger \right)_{II}
\left( V^{({\cal Q}_y)}_R 
m^2_{\tilde{\cal Q}_R} 
{V^{({\cal Q}_y)}_R}^\dagger \right)_{JJ}}}, 
\nonumber \\
(\delta^{({\cal Q}_y)}_{LL})_{IJ} 
&\equiv& \frac{ \left( 
V^{({\cal Q}_y)}_L 
m^2_{\tilde{\cal Q}_L} 
{V^{({\cal Q}_y)}_L}^\dagger \right)_{IJ}}{ 
\sqrt{ 
\left( V^{({\cal Q}_y)}_L 
m^2_{\tilde{\cal Q}_L} 
{V^{({\cal Q}_y)}_L}^\dagger \right)_{II}
\left( V^{({\cal Q}_y)}_L 
m^2_{\tilde{\cal Q}_L} 
{V^{({\cal Q}_y)}_L}^\dagger \right)_{JJ}}}, 
\nonumber \\
(\delta^{({\cal Q}_y)}_{RR})_{IJ} 
&\equiv& \frac{ \left( 
V^{({\cal Q}_y)}_R 
m^2_{\tilde{\cal Q}_R} 
{V^{({\cal Q}_y)}_R}^\dagger \right)_{IJ}}{
\sqrt{ 
\left( V^{({\cal Q}_y)}_R 
m^2_{\tilde{\cal Q}_R} 
{V^{({\cal Q}_y)}_R}^\dagger \right)_{II}
\left( V^{({\cal Q}_y)}_R 
m^2_{\tilde{\cal Q}_R} 
{V^{({\cal Q}_y)}_R}^\dagger \right)_{JJ}}}, 
\nonumber
\end{eqnarray}
where $V^{({\cal Q}_y)}_{L,R}$ 
are given in Eq.~(\ref{eq:mevs}), 
the matrices 
$(a^{({\cal Q}_y)})_{IJ} \equiv 
y^{({\cal Q}_y)}_{IJ \hat{K}=1} 
A^{({\cal Q}_y)}_{IJ \hat{K}=1}$ 
originate from the scalar trilinear couplings, 
and $\hat{K}$ labels the eigenstates defined by Eq.~(\ref{eq:mukl}). 
The superscripts ${\cal Q}_y$ represent ${\cal Q}_y=U,D,N,E$ 
indicating the corresponding subscripts 
$\tilde{\cal Q}_L=\tilde{q},\tilde{q},\tilde{l},\tilde{l}$ and 
$\tilde{\cal Q}_R=\tilde{u},\tilde{d},\tilde{\nu},\tilde{e}$, 
respectively. 

\begin{figure}[t]
\hfill
\includegraphics[width=0.38\linewidth]{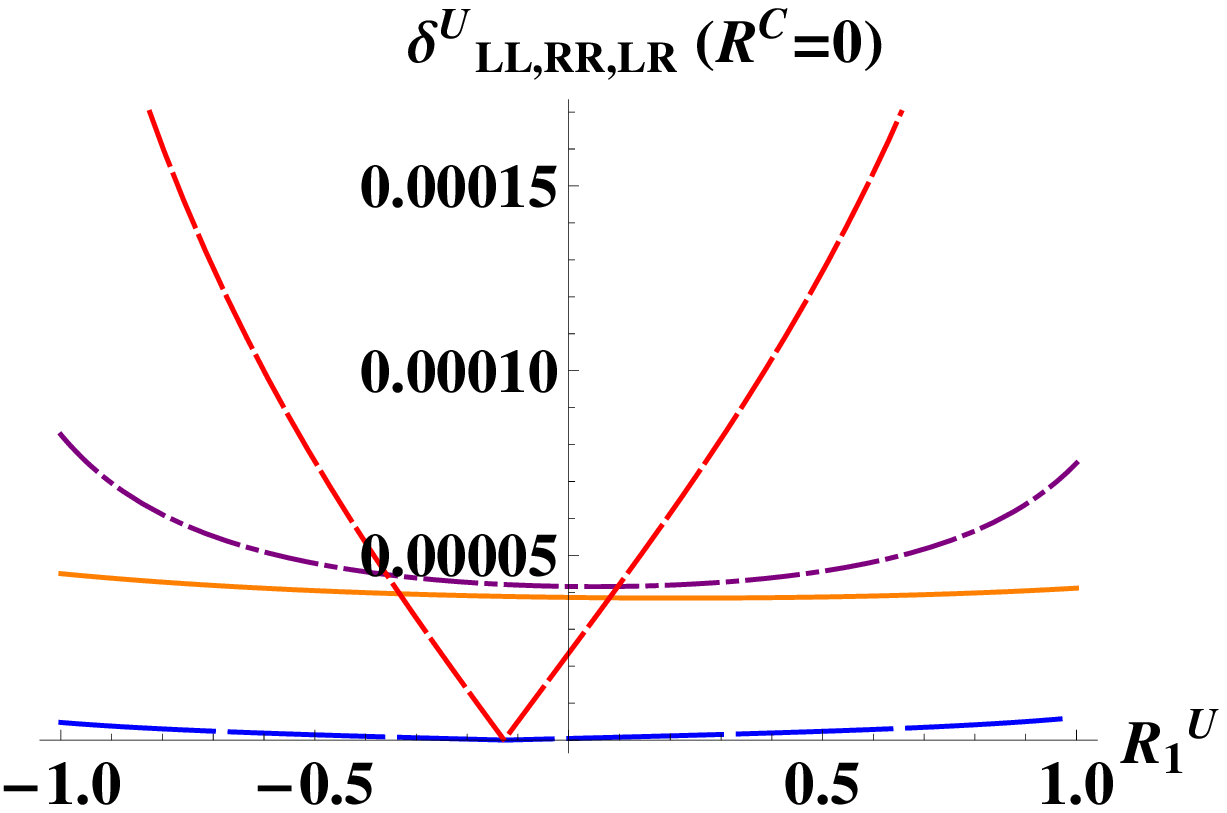}
\hfill
\includegraphics[width=0.38\linewidth]{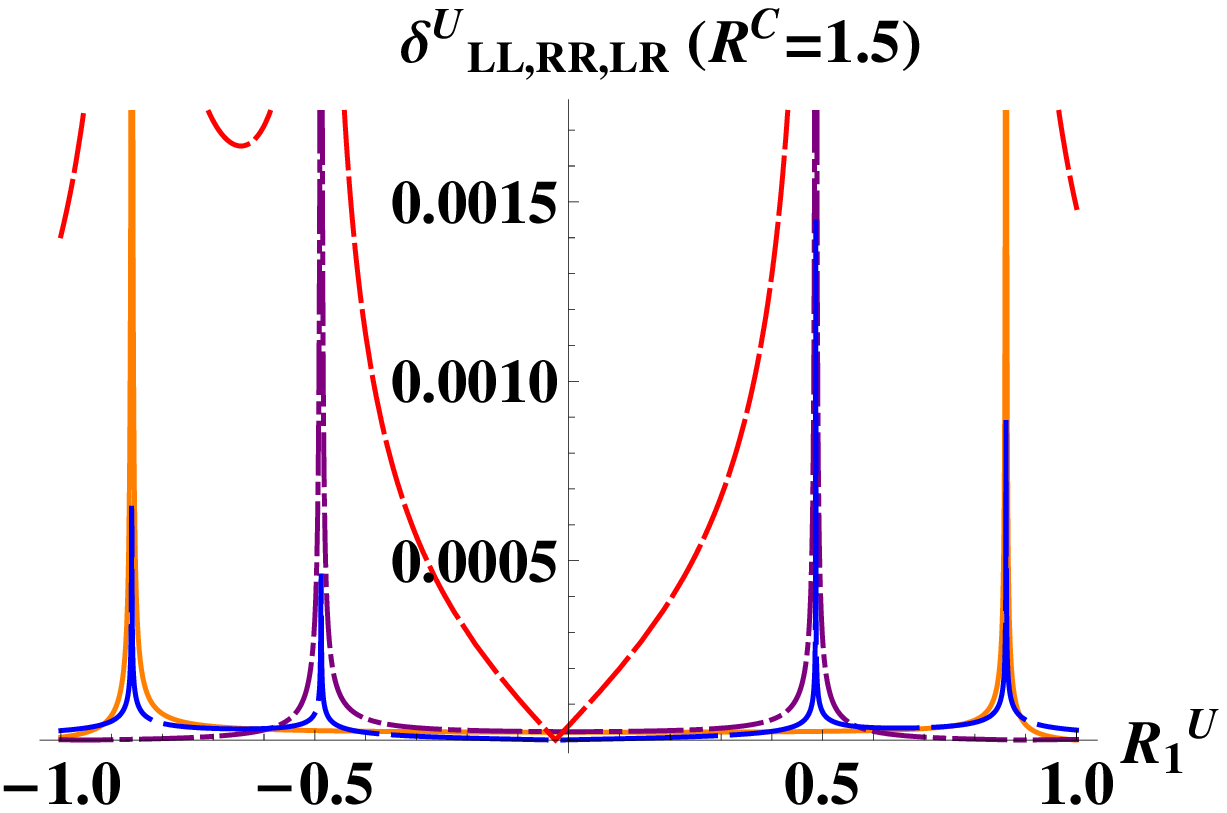}
\hfill
\includegraphics[width=0.20\linewidth]{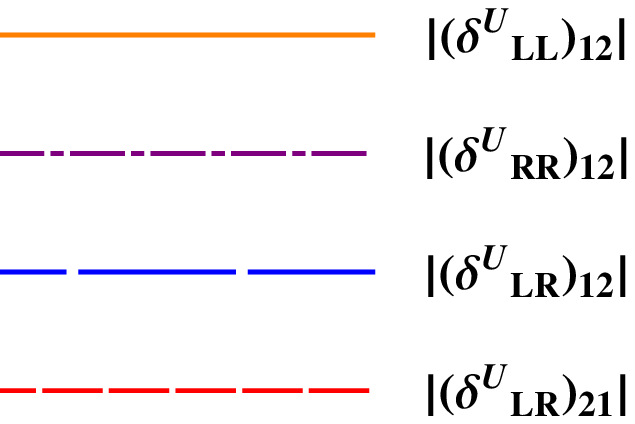}
\caption{The mass insertion parameters 
$(\delta^{(U)}_{LR,LL,RR})_{IJ}$ as a function of 
$R^U_1$ evaluated at the same sample point in the moduli space 
as in Table~\ref{tab:ckm} with the fixed values of 
$R^U_{r \ne 1} = 0.9$, $R^T_r =1$ and $M_{\rm SB}=1$ TeV. }
\label{fig:du}
\end{figure}

\begin{figure}[t]
\hfill
\includegraphics[width=0.38\linewidth]{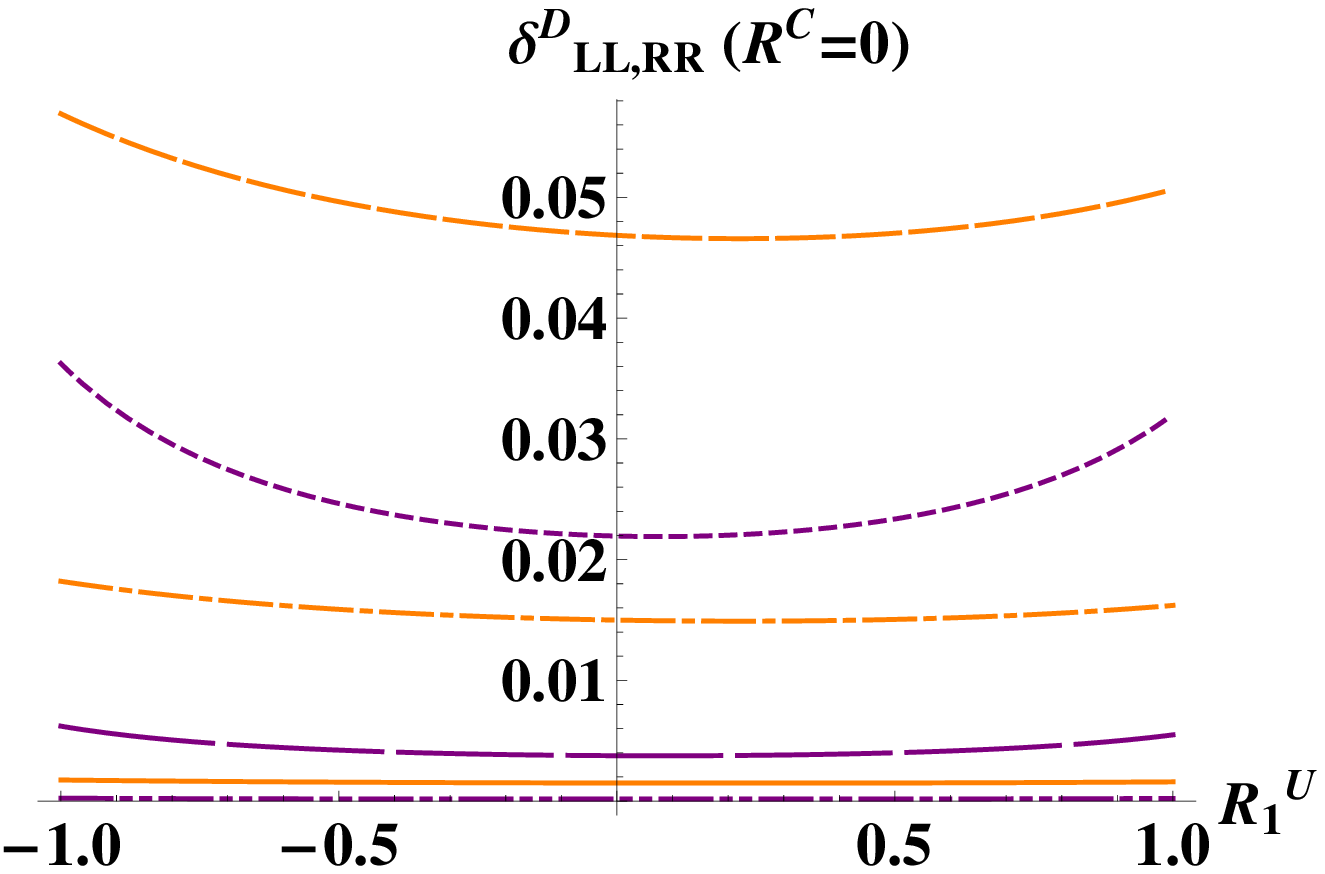}
\hfill
\includegraphics[width=0.38\linewidth]{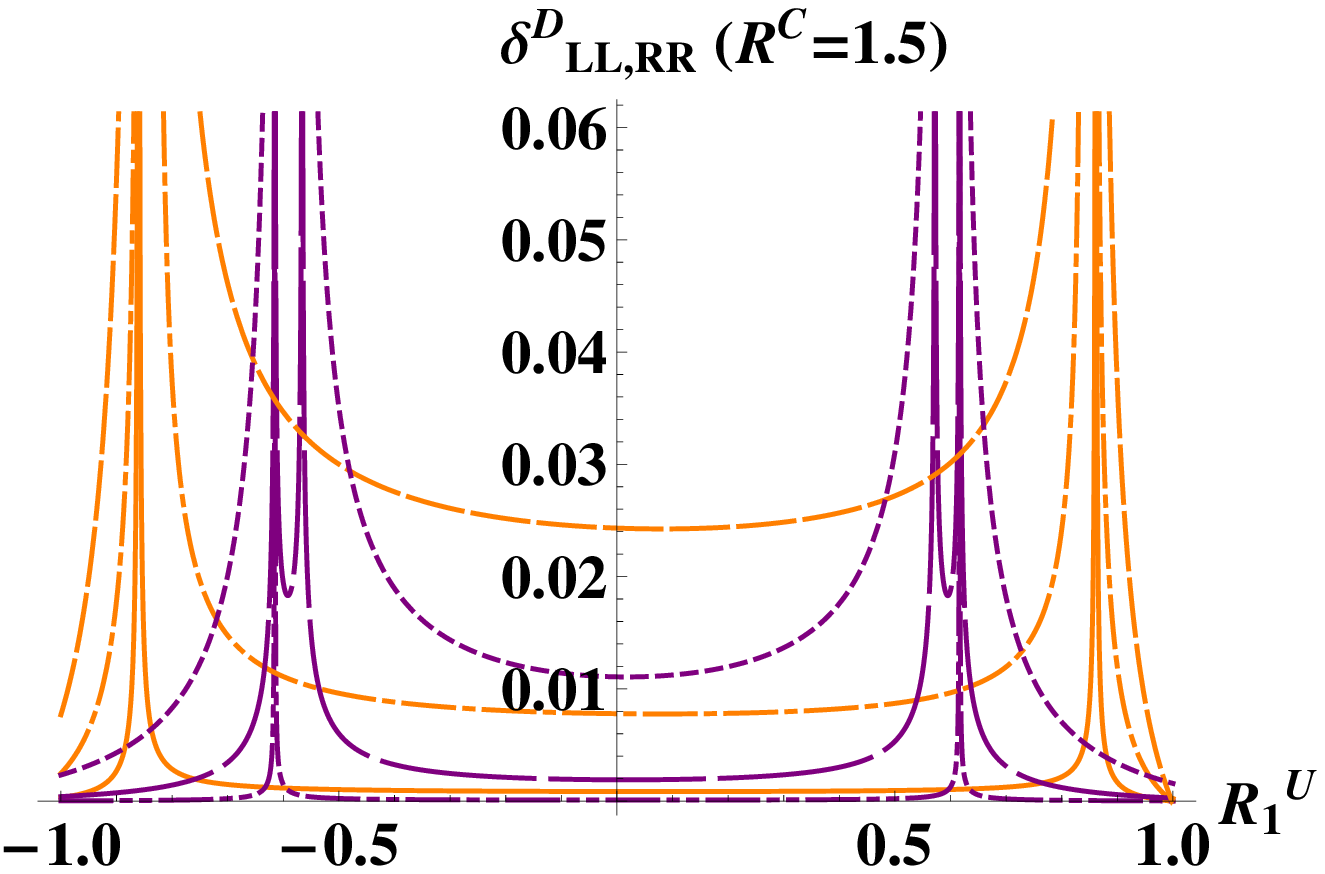}
\hfill
\includegraphics[width=0.20\linewidth]{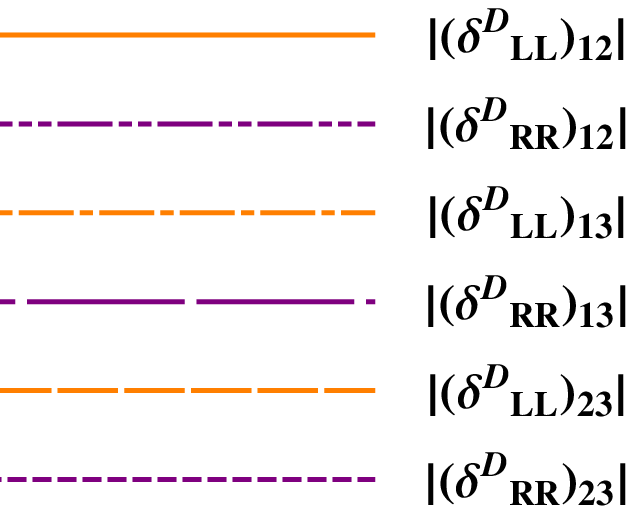}
\\ \ \\
\hfill
\includegraphics[width=0.38\linewidth]{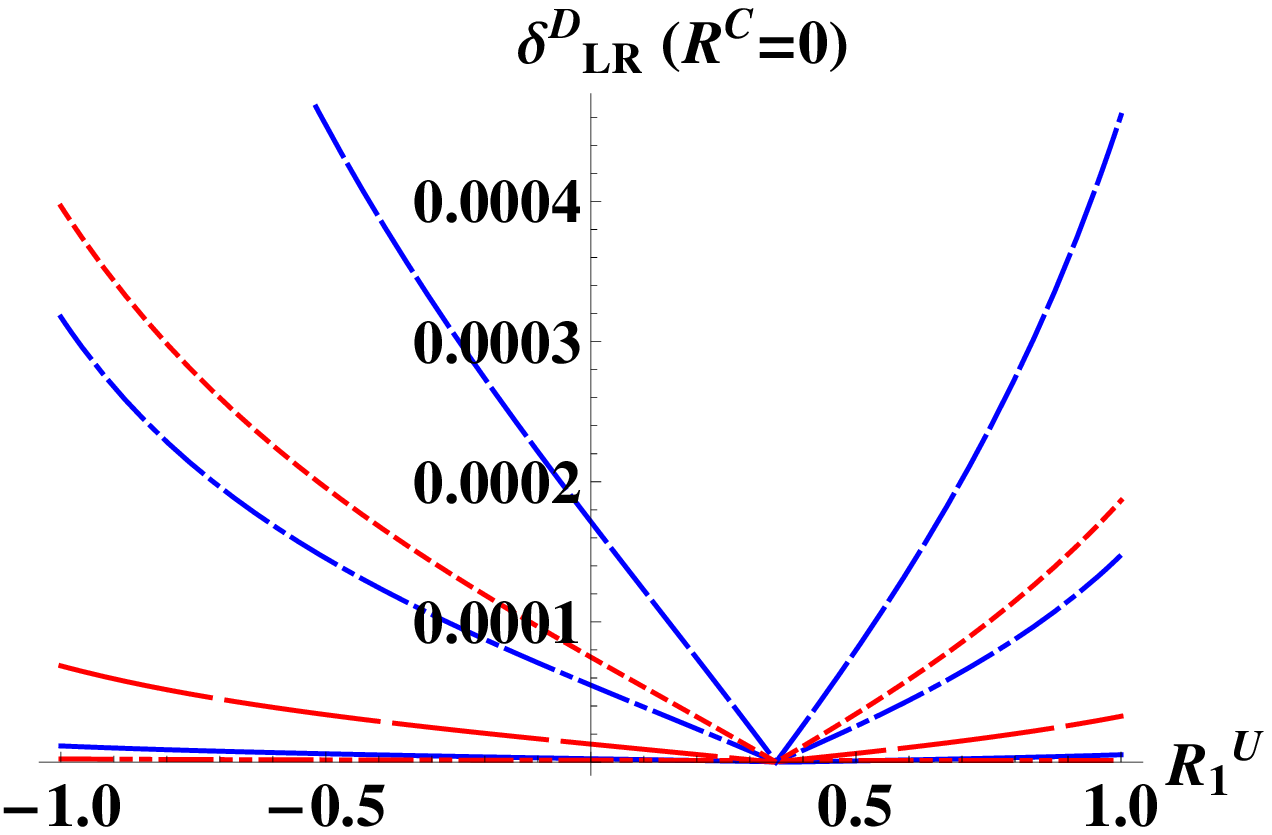}
\hfill
\includegraphics[width=0.38\linewidth]{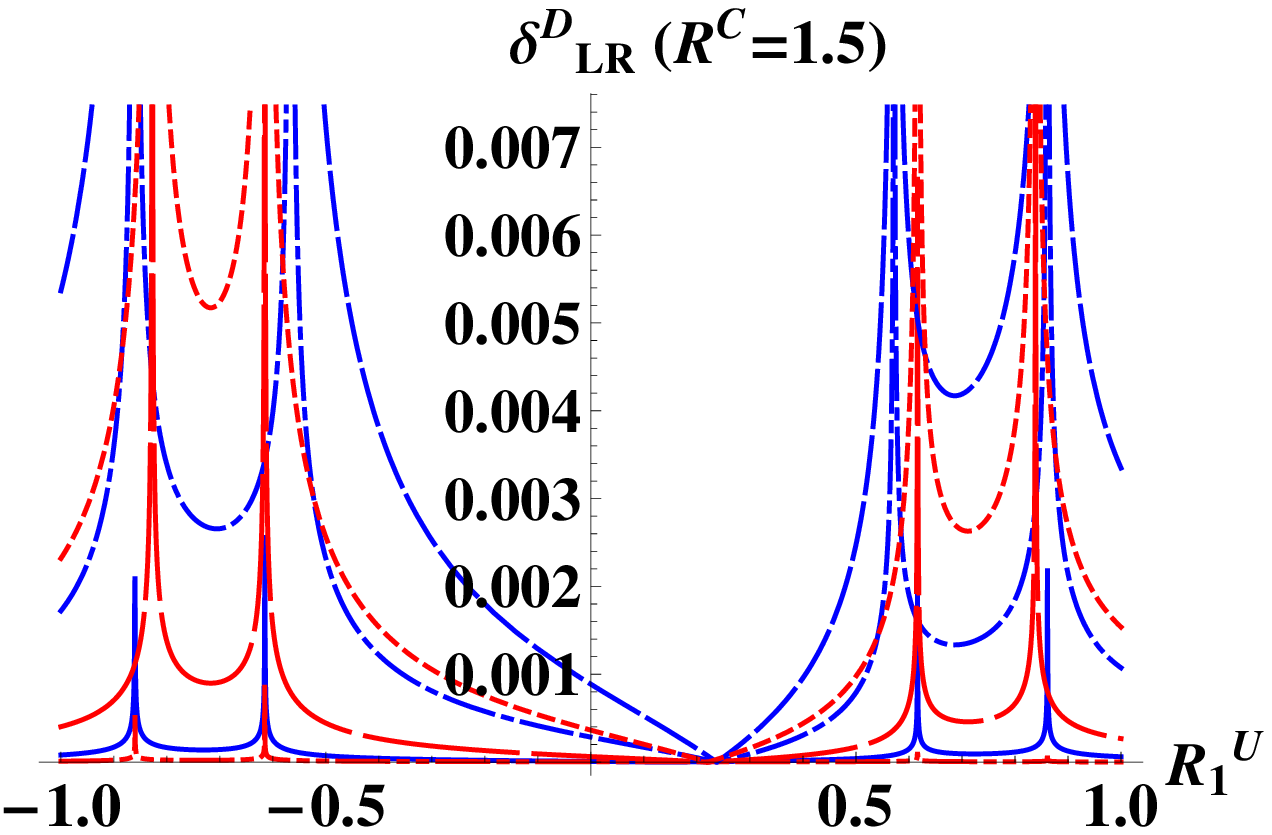}
\hfill
\includegraphics[width=0.20\linewidth]{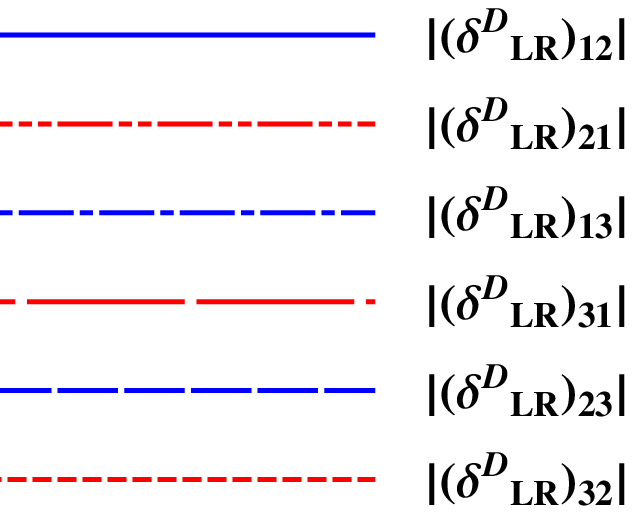}
\caption{The mass insertion parameters 
$(\delta^{(D)}_{LR,LL,RR})_{IJ}$ as a function of 
$R^U_1$ evaluated at the same sample point in the moduli space 
as in Table~\ref{tab:ckm} with the fixed values of 
$R^U_{r \ne 1} = 0.9$, $R^T_r =1$ and $M_{\rm SB}=1$ TeV.}
\label{fig:dd}
\end{figure}

\begin{figure}[t]
\hfill
\includegraphics[width=0.38\linewidth]{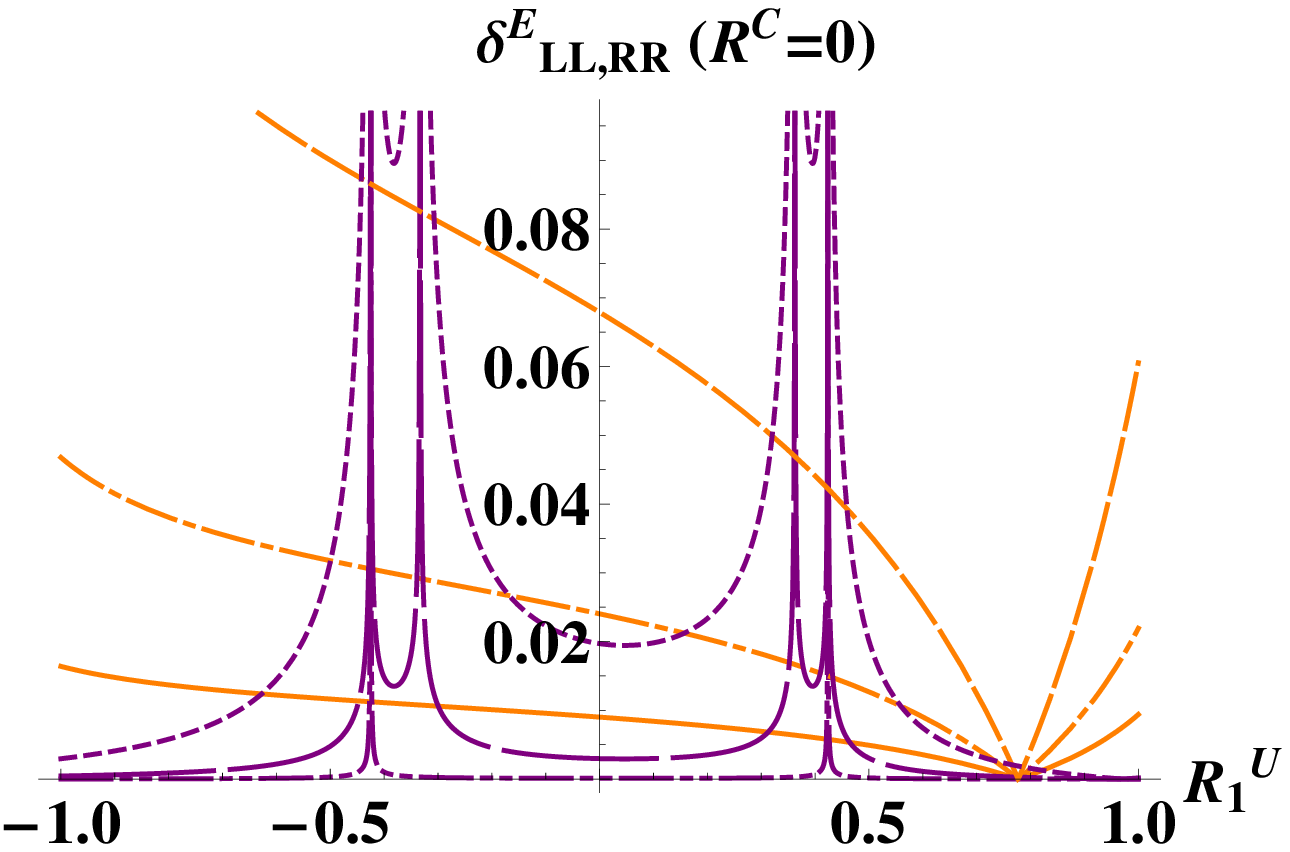}
\hfill
\includegraphics[width=0.38\linewidth]{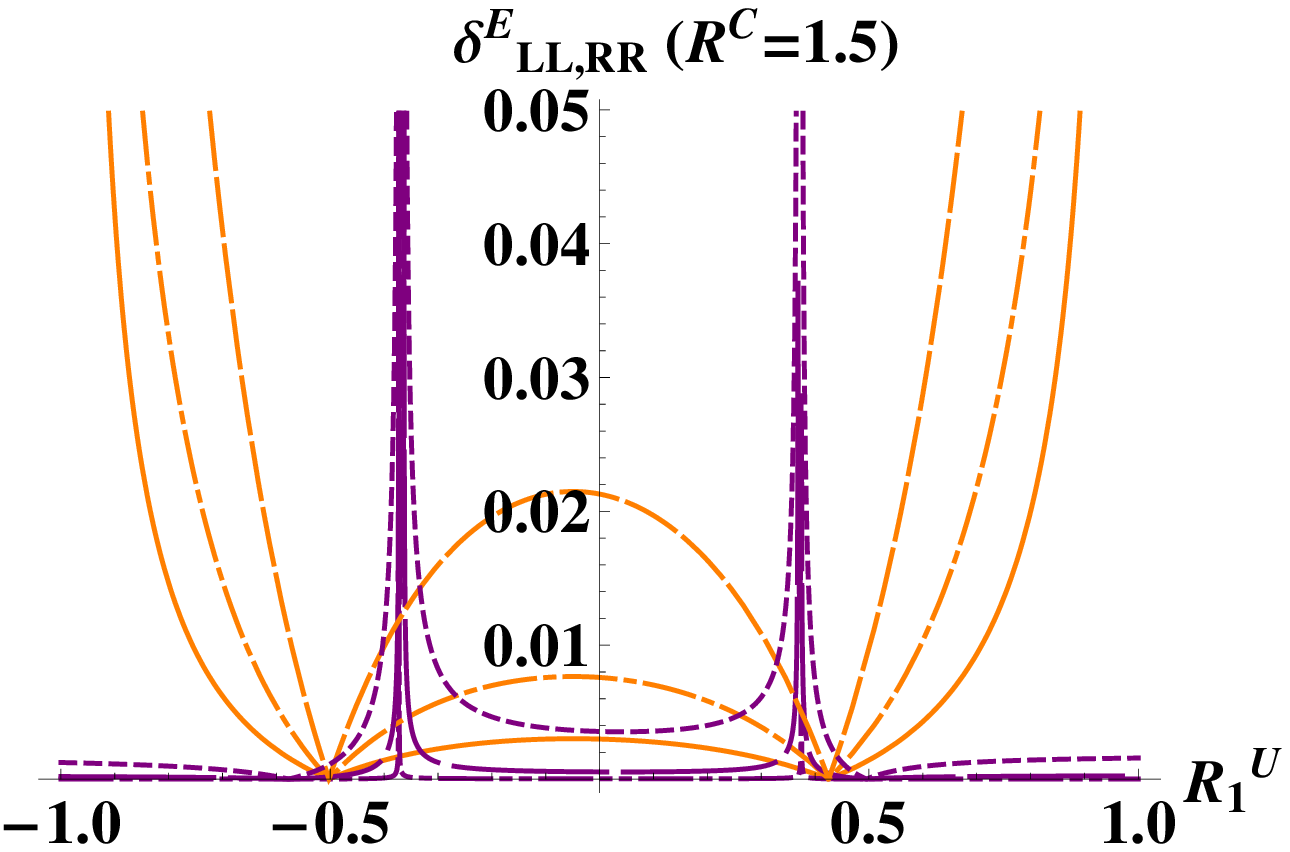}
\hfill
\includegraphics[width=0.20\linewidth]{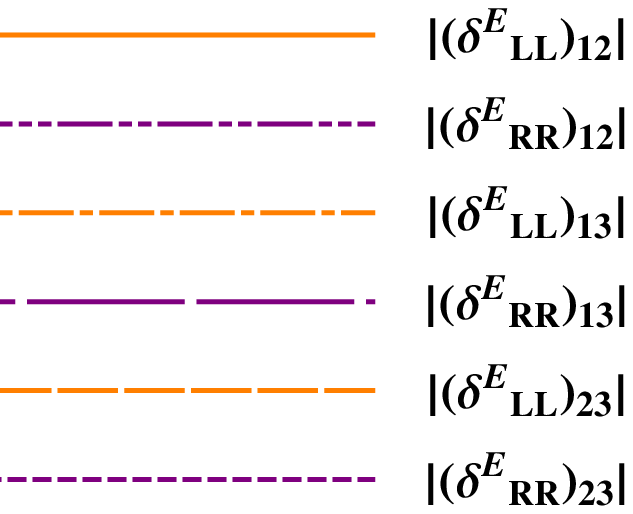}
\caption{The mass insertion parameters 
$(\delta^{(E)}_{LL,RR})_{IJ}$ as a function of $R^U_1$ 
evaluated at the same sample point in the moduli space 
as in Table~\ref{tab:ckm} with the fixed values of 
$R^U_{r \ne 1} = 0.9$, $R^T_r =1$ and $M_{\rm SB}=1$ TeV.}
\label{fig:dellrr}
\end{figure}

\begin{figure}[t]
\hfill
\includegraphics[width=0.38\linewidth]{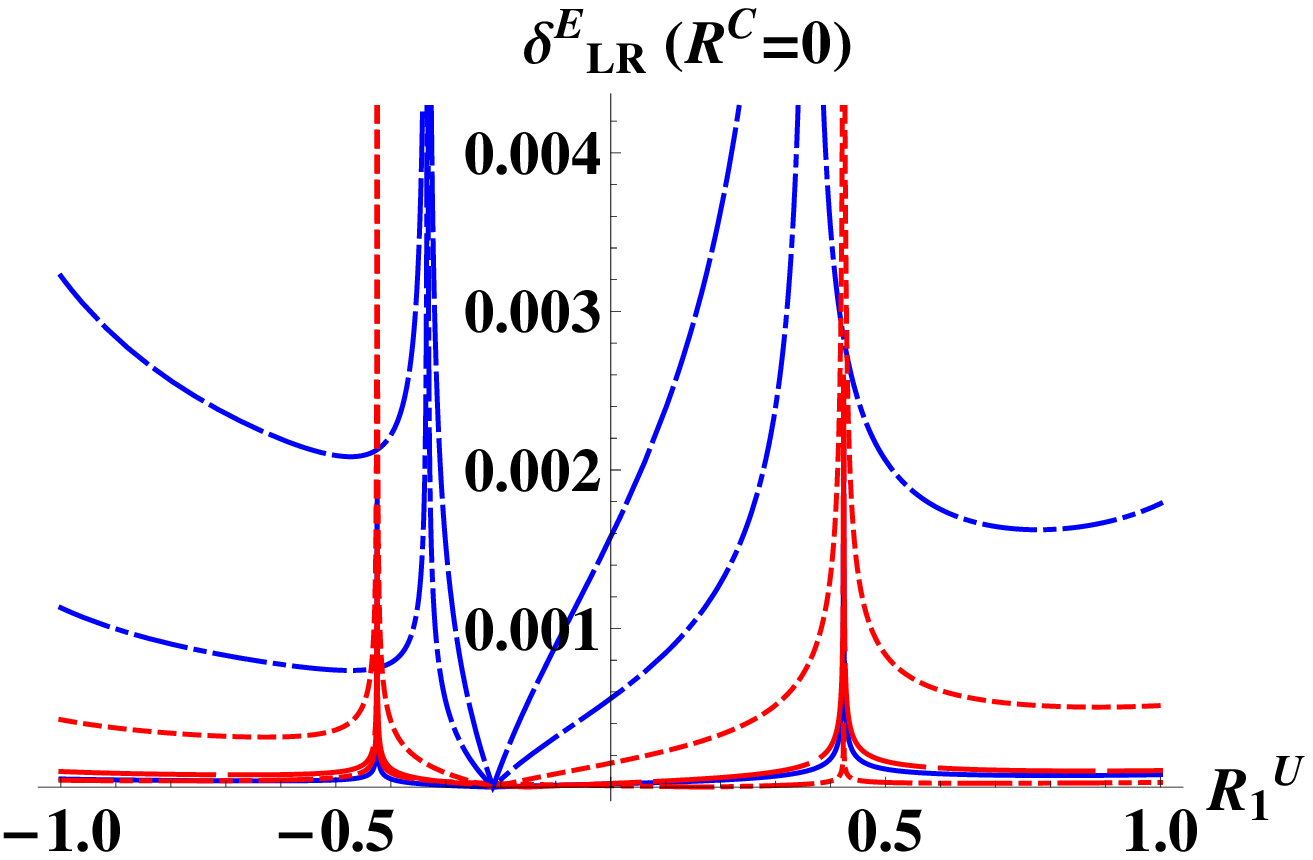}
\hfill
\includegraphics[width=0.38\linewidth]{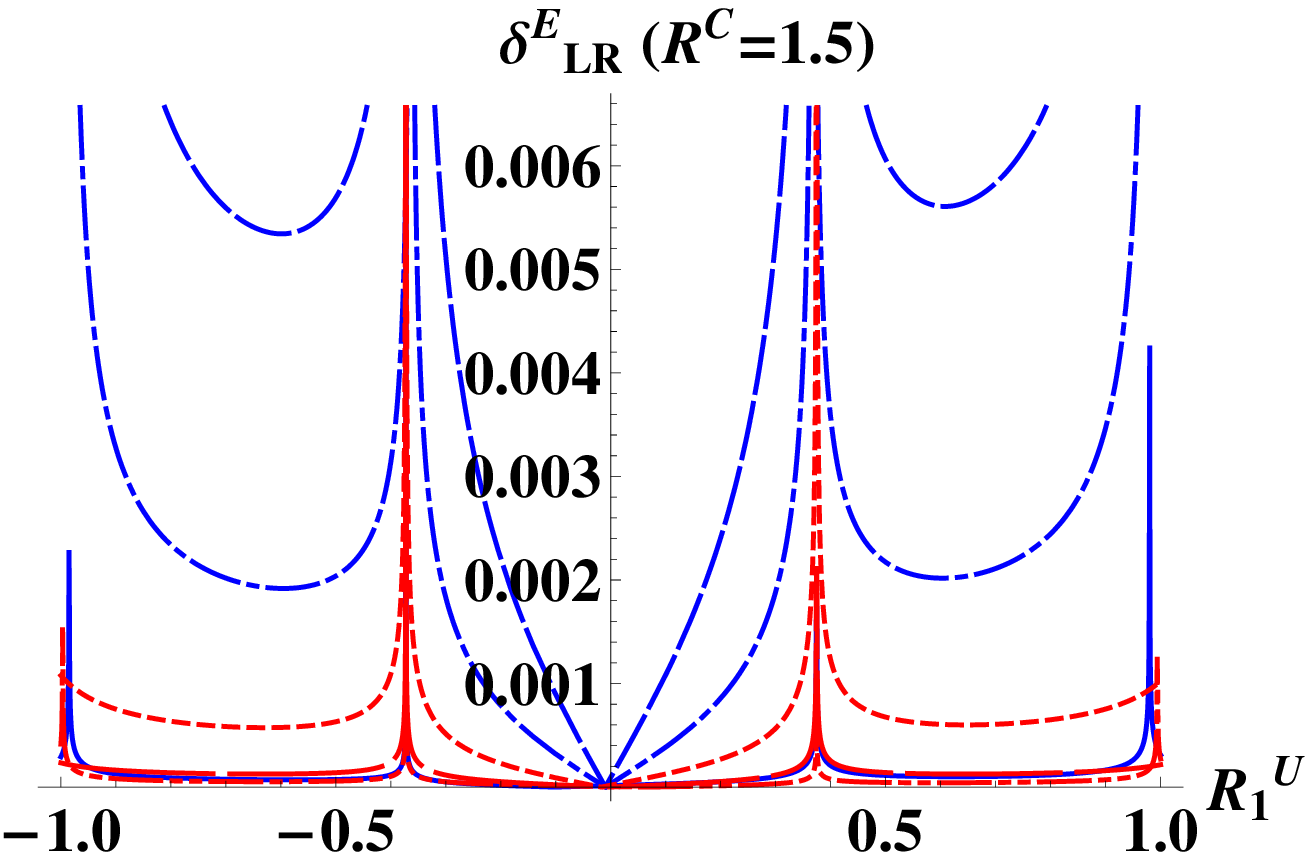}
\hfill
\includegraphics[width=0.20\linewidth]{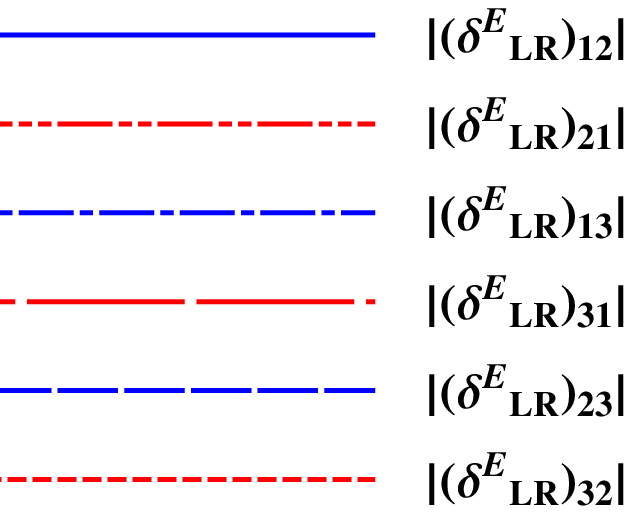}
\\ \ \\
\hfill
\includegraphics[width=0.38\linewidth]{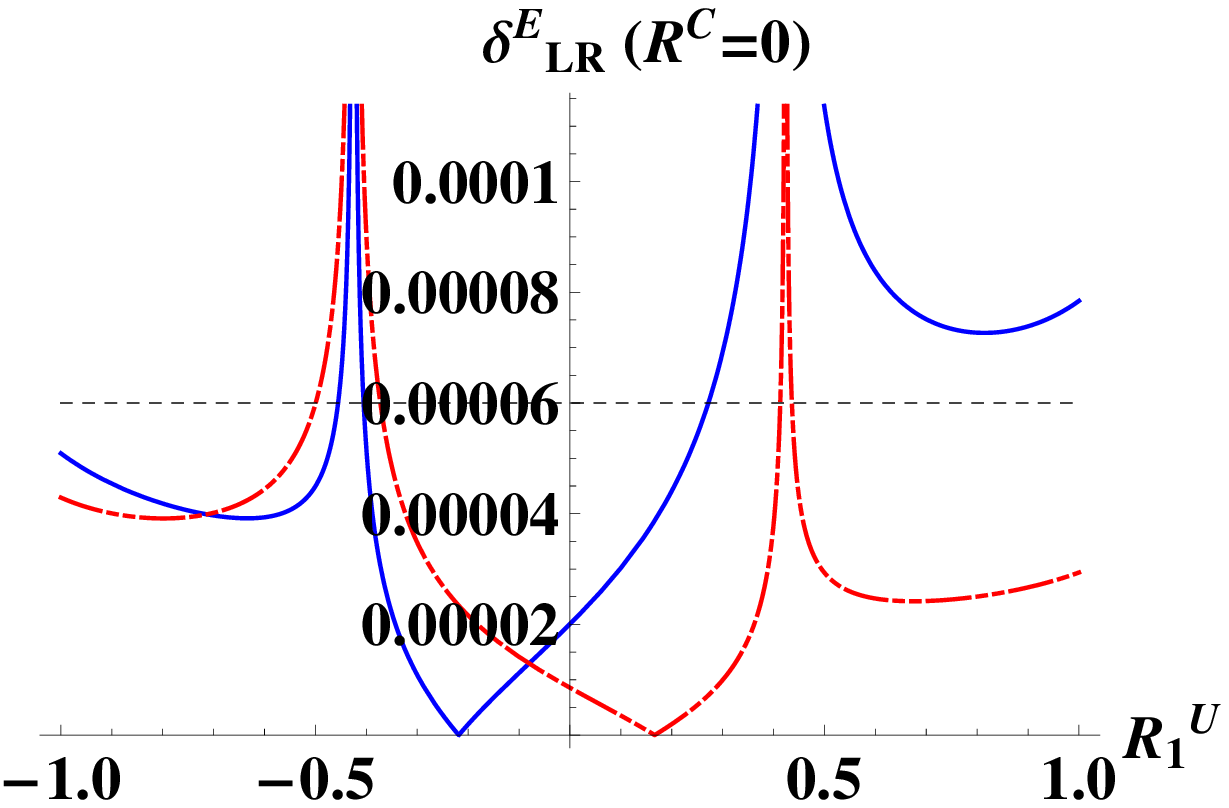}
\hfill
\includegraphics[width=0.38\linewidth]{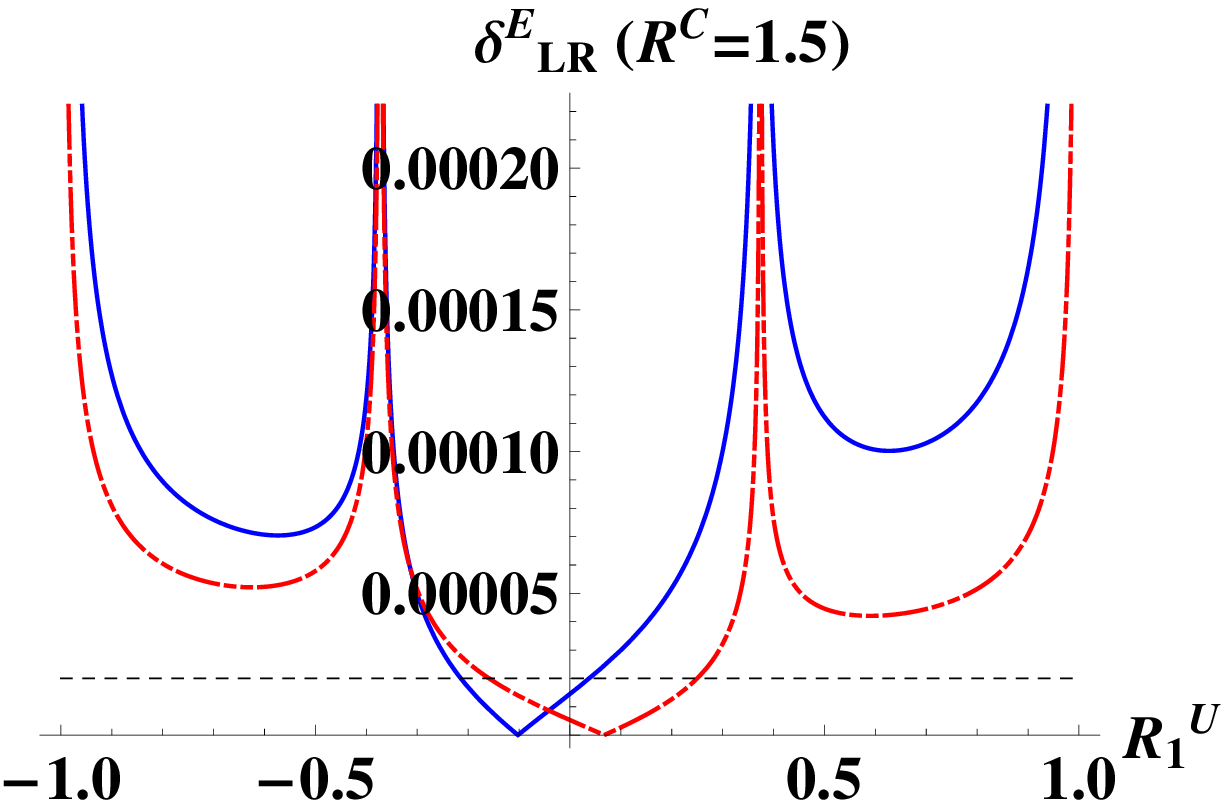}
\hfill
\includegraphics[width=0.20\linewidth]{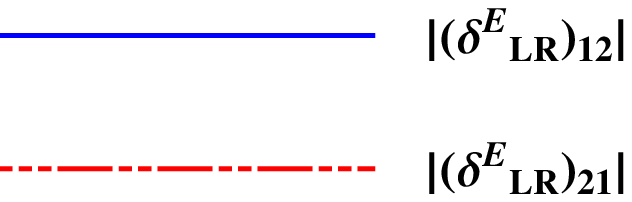}
\caption{The mass insertion parameters 
$(\delta^{(E)}_{LR})_{IJ}$ as a function of $R^U_1$ 
evaluated at the same sample point in the moduli space 
as in Table~\ref{tab:ckm} with the fixed values of 
$R^U_{r \ne 1} = 0.9$, $R^T_r =1$ and $M_{\rm SB}=1$ TeV. 
The horizontal dashed lines in the lower panels represent 
a typical value of the experimental upper bound 
restricting FCNCs that enhances $\mu \to e \gamma$ 
transitions~\cite{Misiak:1997ei}.}
\label{fig:de}
\end{figure}

The $R^U_1$ dependence of the mass insertion parameters is  
shown in Figs.~\ref{fig:du},~\ref{fig:dd},~\ref{fig:dellrr} 
and \ref{fig:de}. 
Each parameter is constrained by various experiments, and 
most of them are free from these constraints in our model 
with $M_{\rm SB} = 1$ TeV. 
However only one of those, the upper bound of 
$(\delta^{(E)}_{LR})_{12,21}$ restricting FCNCs that enhances 
$\mu \to e \gamma$ transitions~\cite{Misiak:1997ei}, 
is very severe as shown in Fig.~\ref{fig:de}. 
From this figure, we find that the value of $F^U_1$ 
is severely restricted, that is, the amount of supersymmetry 
breaking mediated by the complex-structure (shape) modulus 
of the first torus, $U_1$, must be extremely small. 
That is expected from the fact that only the $U_1$ 
distinguishes the flavors (the differences between the 
wavefunction profiles of chiral matter fields on the 
first torus) as can be seen in the expressions of Yukawa 
couplings~(\ref{eq:yukawa}). 
On the other hand, all the other moduli $S$, $T_r$, $U_{r \ne 1}$ 
can mediate sizable supersymmetry breaking without conflicting 
with the experimental data concerning supersymmetric flavor 
violations. These flavor violations become smaller for 
larger values of $M_{\rm SB}$ due to the decoupling 
effect of the superparticles. 

\subsection{A typical superparticle spectrum}

\begin{figure}[t]
\begin{center}
\includegraphics[width=0.45\linewidth]{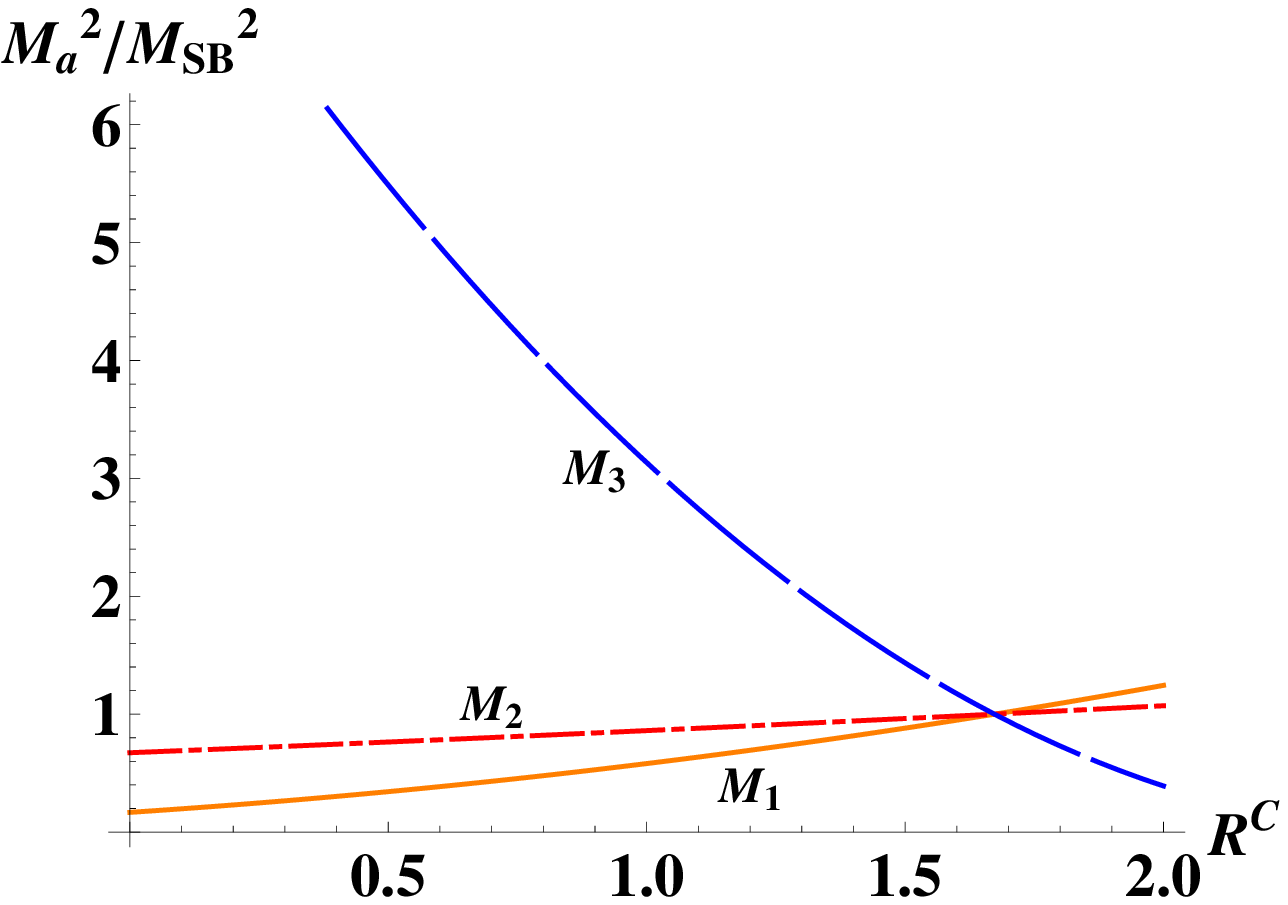}\\
\includegraphics[width=0.45\linewidth]{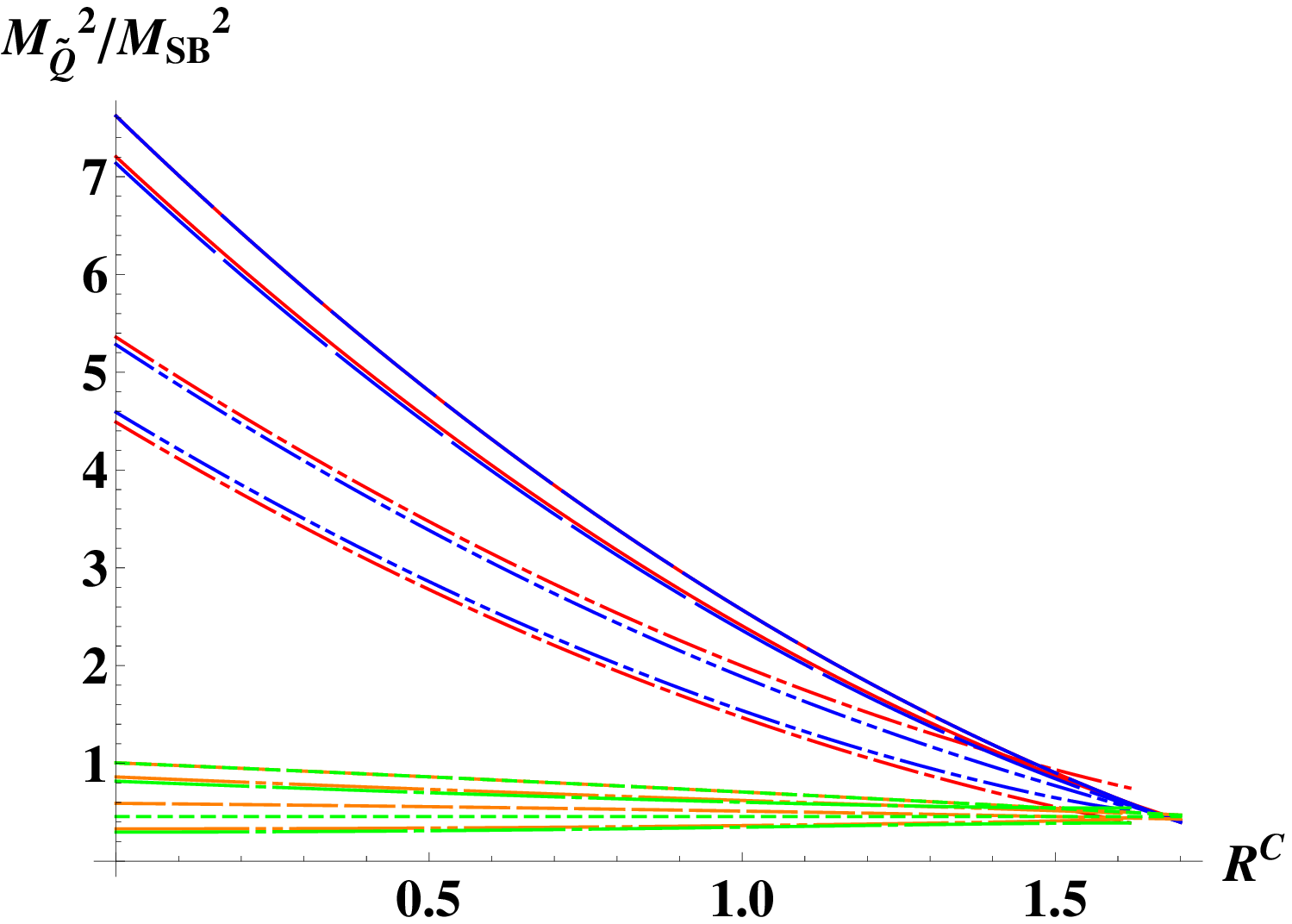}
\includegraphics[width=0.2\linewidth]{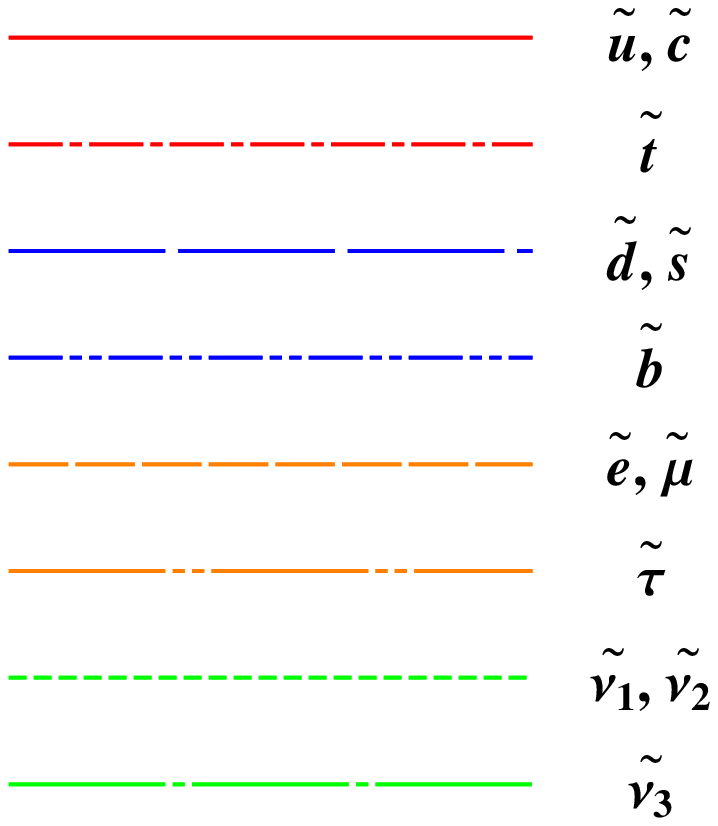}
\hfill
\end{center}
\caption{
The masses of the gauginos (the top panel) 
and of the sfermions (the bottom panel) 
at the EW scale as functions of $R^C$ 
evaluated at the same sample point in the moduli space 
as in Table~\ref{tab:ckm} with fixed values of 
$R^U_1 = -0.05$, $R^U_{r \ne 1} = 0.9$, $R^T_r =1$ 
and $M_{\rm SB}=1$ TeV.}
\label{fig:gauge}
\end{figure}

We show a typical superparticle spectrum at the EW scale 
by varying $R^C$ with fixed values of $M_{\rm SB}$, 
$R^U_r$, $R^T_r$ and $\tan \beta$ in Fig.~\ref{fig:gauge}. 
The supersymmetry breaking scale is again fixed as $M_{\rm SB}=1$ TeV. 
Because the value of $R^U_{r=1}$ is severely constrained as shown 
in Fig.~\ref{fig:de}, an allowed small value $R^U_1=-0.05$ is chosen, 
while the ratio $R^U_{r\neq1}$ and $R^T_r$ does not affect the 
spectrum so much and then $R^U_{r\neq1}=0.9$ and $R^T_r=1$ ($r=1,2,3$) 
are adopted here. As mentioned previously, the $\mu$-parameter 
is fixed in such a way that the EW symmetry is broken successfully 
yielding the observed masses of $W$ and $Z$ bosons. 
Curves describing some soft scalar masses in Fig.~\ref{fig:gauge} 
are terminated at $R^C \sim 1.6$, because the EW symmetry is not 
broken successfully with $R^C \gtrsim 1.6$. 

A mediation mechanism of supersymmetry breaking, which is 
a sizable mixture of the modulus and the anomaly mediation, namely 
$R^C \sim {\cal O}(1)$, is called the mirage mediation~\cite{Choi:2005uz}. 
Especially, the mass spectrum with $R^C \sim 1.6$ in our model, 
where the gaugino masses and scalar masses respectively degenerate 
at the TeV scale, resembles that of the TeV scale mirage mediation 
model~\cite{Choi:2005hd}. It is pointed out in this model that 
the notorious fine-tuning between supersymmetric and supersymmetry 
breaking parameters in the MSSM is dramatically ameliorated. 

As for the lightest superparticle in the above spectrum, we find 
that it is a neutralino.  The eigenvalues of the neutralino and the 
chargino masses measured in the unit of GeV are listed in the 
following table.

\begin{center}
\begin{tabular}{|c||c|c|} \hline
 & Neutralino & Chargino \\ \hline
$R^C=0$ & 
$(\,1824, \,1822, \,819, \,409\,)$ & 
$(\,1822, \,820\,)$ 
\\ \hline
$R^C=1.5$ & 
$(\,986, \,941, \,395, \,388\,)$ & 
$(\,982, \,392\,)$ 
\\ \hline
\end{tabular}
\end{center}

\begin{figure}[thb]
\begin{center}
\includegraphics[width=0.45\linewidth]{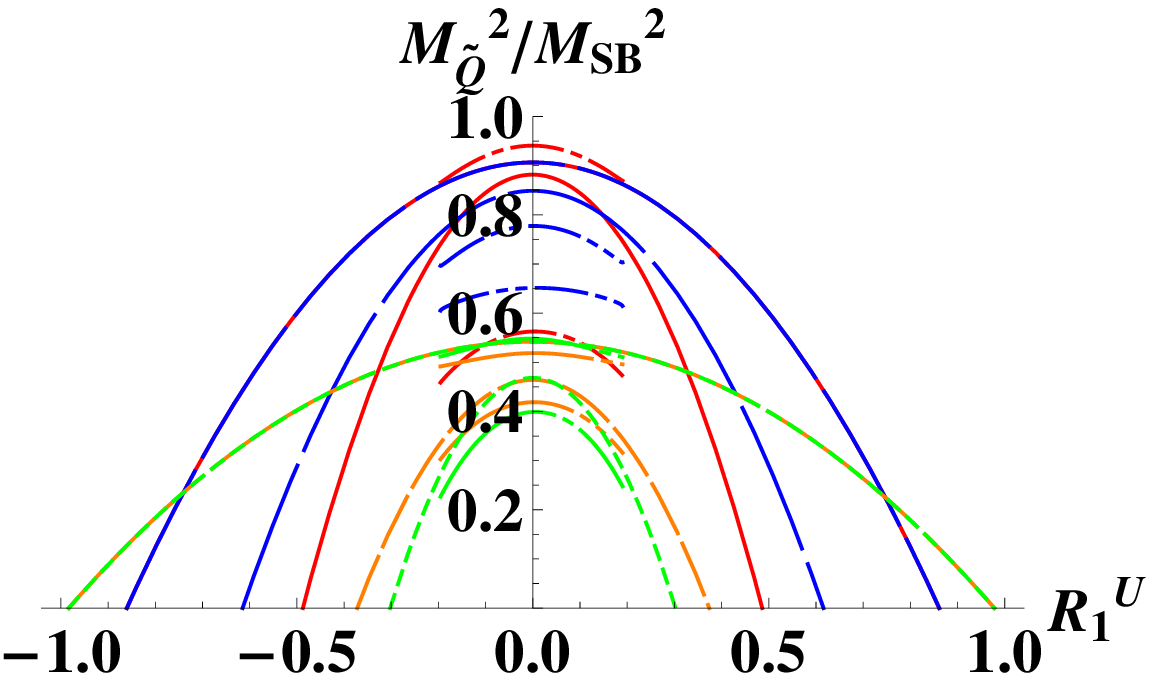}
\includegraphics[width=0.2\linewidth]{spectrum_legend.eps}
\hfill
\end{center}
\caption{
The masses of the sfermions 
at the EW scale as functions of $R^U_1$ 
evaluated at the same sample point in the moduli space 
as in Table~\ref{tab:ckm} with fixed values of 
$R^C = 1.5$, $R^U_{r \ne 1} = 0.9$, $R^T_r =1$ 
and $M_{\rm SB}=1$ TeV.}
\label{fig:spec}
\end{figure}

So far, we have considered the scenario with a low-energy supersymmetry breaking, and selected a small value $R^U_1=-0.05$ to be consistent with 
the experimental data concerning supersymmetric flavor violations. 
If we consider the case with larger values of $M_{\rm SB}$, 
with which the flavor violations 
become smaller, the values of $R^U_1$ can reside in much wider region. 
However, there are other two factors restricting 
the values of $R^U_1$ besides those from FCNCs. 
One is related to the success of the EW symmetry breaking, and the other is 
related to obtaining non-tachyonic masses. 
We show the $R^U_1$ dependence of the masses of sfermions 
with $R^C = 1.5$ and $M_{\rm SB}= 1$ TeV in Fig~\ref{fig:spec}. 
In the figure, some curves are terminated at $R^U_1\sim \pm 0.2$ 
because the EW symmetry is not broken successfully for $|R^U_1| \gtrsim 0.2$, 
as the situation in Fig~\ref{fig:gauge}. We find that $R^U_1$ has to be in the range 
$\left|R^U_1\right| < 0.2$, where we obtain non-tachyonic masses.
With other values of $R^C$ and $M_{\rm SB}$, it is possible that the 
non-tachyon condition is more severe than the other. 
In some typical cases with 
($R^C, M_{\rm SB}/$ TeV) = (0, 1), (0, 10),(1.5, 1) and (1.5, 10), 
we also find that the allowed region of the ratio $R^U_1$, 
where the EW symmetry is broken successfully and 
non-tachyonic masses are obtained, 
is roughly $\left|R^U_1\right| < 0.2$. 
That has to be in mind, especially when one considers larger values of $M_{\rm SB}$. 
 
Finally, we comment on the Higgs sector. 
In our model, there are some possibilities to obtain the 
mass of the lightest CP-even Higgs boson $m_h \sim 125$ GeV, 
which is indicated from the recent observations at the 
LHC~\cite{:2012gk}. 
First of all, as is well known, we can easily realize 
$m_h \sim 125$ GeV with $M_{\rm SB} \sim 10$ TeV. 
The supersymmetric flavor violations are much smaller 
in this case than those we have studied above for 
$M_{\rm SB} = 1$ TeV, and the bound on $R^U_1$ from the FCNCs 
disappears. Then, in this case $|R^U_1|<0.2$ is suggested. 

The second possibility is that, we can consider the next-to 
MSSM in some extensions of our model where $m_h \sim 125$ GeV 
could be realized with a low scale supersymmetry breaking 
$M_{\rm SB} \sim 1$ TeV (see for review
e.g. Ref.~\cite{Ellwanger:2009dp}). 
In this case, the supersymmetric 
flavor violations and the superparticle spectrum estimated 
above can be applied straightforwardly. Some analyses of 
such an extended Higgs sector in the TeV-scale mirage mediation models 
are performed in Ref.~\cite{Asano:2012sv}. 

Besides these two, there is one more interesting possibility. 
Although we have worked on the 10D SYM theory in this paper, 
it would be straightforward to extend our model to SYM theories 
in a lower-than-ten dimensional spacetime, or even to the mixture 
of SYM theories with a different dimensionality. For example, 
in type IIB orientifolds, our model will be adopted not only to the 
magnetized D9 branes (a class of which is T-dual to intersecting 
D6 branes in IIA side), but also to the D5-D9~\cite{DiVecchia:2011mf} 
and the D3-D7 brane configurations with magnetic fluxes 
in the extra dimensions. An interesting possibility is that the 
$SU(3)_C$ and the $SU(2)_L$ gauge groups of the MSSM originate 
from different branes with a different dimensionality, and then 
the moduli-dependence of the gauge kinetic functions are different 
by the gauge groups, that can cause nonuniversal gaugino masses at 
the tree level in the effective supergravity action. 
The situation may allow $m_h \sim 125$ GeV just within the 
MSSM with a low scale supersymmetry breaking without 
a severe fine-tuning~\cite{Abe:2007kf}. 

Even in this case the same flavor structures in the MSSM sector 
would be realized as those in the 10D model presented in this paper, 
if these two branes share a single magnetized torus $T^2$ of the 
same structure as the first torus ($r=1)$ in our 10D model. 
Furthermore, the mixed brane configurations may allow an introduction 
of the supersymmetry-breaking branes sequestered from the visible sector, 
which coincide with the flavor structure derived in this paper. 
The model building based on such mixed brane configurations will 
be reported in separate papers~\cite{WIP}.

\section{Conclusions and discussions}
\label{sec:conc}

We have constructed a three-generation model of quark and lepton 
chiral superfields based on a toroidal compactification of the 
10D SYM theory with certain magnetic fluxes in extra dimensions 
preserving a 4D ${\cal N}=1$ supersymmetry. 
The low-energy effective theory contains the MSSM particle contents, 
where the numbers of chiral generations are determined by the numbers 
of the fluxes they feel, and the most massless exotics can be 
projected out by a combinatory effect of the magnetic fluxes 
and a certain orbifold projection. 

We find that a semi-realistic pattern of the quark and the charged 
lepton masses and the CKM mixings is realized at a certain sample 
point in the (tree-level) moduli space of the 10D SYM theory, where the 
VEVs of the six Higgs doublets and of the geometric moduli as well as 
the Wilson-line parameters take reasonable numerical values without 
any hierarchies. In addition, it has been shown that a semi-realistic 
pattern of the neutrino masses and the PMNS mixings can be achieved 
at the same point of the moduli space, if we assume the existence 
of certain effective superpotential terms~(\ref{eq:wmm}), those would 
be induced by nonperturbative effects and/or higher-order corrections. 
We have assumed the existence of such nonperturbative effects 
and/or higher-order corrections making the remaining massless 
exotics heavy enough and also generating effectively the neutrino 
Majorana mass term~(\ref{eq:wmm}) as well as the mu-term~(\ref{eq:mu}). 
Further studies are required to find the concrete origin of these effects. 

Thanks to the systematic way of the dimensional reduction in a 
4D ${\cal N}=1$ superspace proposed in Ref.~\cite{Abe:2012ya}, 
the soft supersymmetry breaking parameters induced by the 
moduli-mediated supersymmetry breaking are calculated explicitly. 
Because the flavor structures of our model are essentially determined 
by the localized wavefunctions of the chiral zero-modes, the 4D 
effective theory possesses flavor dependent holomorphic Yukawa 
couplings and flavor independent K\"ahler metrics for the MSSM 
matter fields. Under the assumption that the moduli-mediated 
low scale supersymmetry breaking dominates the soft supersymmetry 
breaking terms in the MSSM, we estimated the size of supersymmetric 
flavor violations by analyzing the mass insertion parameters 
governing various FCNCs, scanning supersymmetry breaking order 
parameters mediated by the dilaton, the geometric moduli and 
the compensator chiral superfields in the 4D ${\cal N}=1$ 
effective supergravity. 

The most stringent bound comes from the $\mu \to e \gamma$ on 
the size of the $F$-term in the chiral multiplet of the complex 
structure modulus of the first torus where the SM flavor structure 
is generated via the wavefunction localization. The result provides 
a strong insight into the mechanism of moduli stabilization in our model. 
For instance, a mechanism of the moduli stabilization proposed by 
Ref.~\cite{Kachru:2003aw} would be suitable, that predicts vanishing 
$F$-terms of the complex structure moduli~\cite{Choi:2005ge,Choi:2004sx} 
at the leading order. Therefore, it would be interesting to study a 
mechanism of the moduli stabilization and the supersymmetry breaking 
at a Minkowski minimum~\cite{Kachru:2003aw} by minimizing the moduli 
and the hidden-sector potential generated by some combinations~\cite{Dudas:2006gr} 
of nonperturbative effects and a dynamical supersymmetry 
breaking~\cite{Intriligator:2006dd}. 

In our model, there are some possibilities to realize the mass of the 
lightest CP-even Higgs boson to be consistent with the recent observations 
at the LHC~\cite{:2012gk}. As mentioned at the end of Sec.~\ref{sec:pheno}, 
especially, it is very interesting to consider the D5-D9~\cite{DiVecchia:2011mf} 
and the D3-D7 brane configurations with magnetic fluxes in the extra dimensions. 
With such brane configurations, we will be able to build more realistic 
models in which we can study concretely the Higgs sector as well as the supersymmetry-breaking sector, the mechanism of moduli stabilization 
and so on. Even in this case the same flavor structures would be realized 
as those in the 10D model presented in this paper, if these two branes 
share a single magnetized torus $T^2$ of the same structure as the first 
torus ($r=1)$ in our 10D model. The model building based on such mixed 
brane configurations will be reported in separate papers~\cite{WIP}.

We have studied on the tree-level 4D effective theory of massless
modes.
Recently, massive modes were studied in Ref.~\cite{Hamada:2012wj}.
They may have phenomenologically important effects on 
4D effective theory.
For example, the K\"ahler potential, superpotential and 
gauge kinetic functions would have threshold corrections due to 
massive modes and such corrections may affect the soft 
supersymmetry breaking terms.
Thus, it is important to study such effects, 
although that is beyond our scope of this paper.

\subsection*{Acknowledgement}
The authors would like to thank T.~Higaki and Y.~Nomoto 
for useful discussions. 
The work of H.~A. was supported by the Waseda University 
Grant for Special Research Projects No.~2012B-151. 
The work of T.~K. is supported in part by  the Grant-in-Aid for 
the Global COE Program ``The Next Generation of Physics, 
Spun from Universality and Emergence'' from the Ministry of 
Education, Culture, Sports, Science and Technology of Japan. 
The work of H.~O. is supported by the JSPS Grant-in-Aid for 
Scientific Research (S) No.~22224003.

\appendix

\section{K\"ahler metrics and holomorphic Yukawa couplings}
\label{app:kmhy}

The K\"ahler potential~$K$, the superpotential~$W$ and 
the gauge kinetic functions $f_a$ in the 4D effective 
supergravity action of our model are derived as~\cite{Abe:2012ya} 
\begin{eqnarray}
K 
&=& K^{(0)}(\bar\Phi^{\bar{m}},\Phi^m) 
+Z^{({\cal Q})}_{\bar{\cal I}{\cal J}}(\bar\Phi^{\bar{m}},\Phi^m) 
\bar{\cal Q}^{\bar{\cal I}} {\cal Q}^{\cal J} 
\nonumber \\ 
&=& K^{(0)}(\bar\Phi^{\bar{m}},\Phi^m) 
\nonumber \\ &&
+Z^{(Q)}_{\bar{I}J}(\bar\Phi^{\bar{m}},\Phi^m) 
\bar{Q}^{\bar{I}} Q^J 
+Z^{(U)}_{\bar{I}J}(\bar\Phi^{\bar{m}},\Phi^m) 
\bar{U}^{\bar{I}} U^J 
+Z^{(D)}_{\bar{I}J}(\bar\Phi^{\bar{m}},\Phi^m) 
\bar{D}^{\bar{I}} D^J 
\nonumber \\ && 
+Z^{(L)}_{\bar{I}J}(\bar\Phi^{\bar{m}},\Phi^m) 
\bar{L}^{\bar{I}} L^J 
+Z^{(N)}_{\bar{I}J}(\bar\Phi^{\bar{m}},\Phi^m) 
\bar{N}^{\bar{I}} N^J 
+Z^{(E)}_{\bar{I}J}(\bar\Phi^{\bar{m}},\Phi^m) 
\bar{E}^{\bar{I}} E^J 
\nonumber \\ && 
+Z^{(H_u)}_{\bar{K}L}(\bar\Phi^{\bar{m}},\Phi^m) 
\bar{H}_u^{\bar{K}} H_u^L 
+Z^{(H_d)}_{\bar{K}L}(\bar\Phi^{\bar{m}},\Phi^m) 
\bar{H}_d^{\bar{K}} H_d^L, 
\nonumber \\*[5pt] 
W 
&=& \lambda^{({\cal Q})}_{{\cal I}{\cal J}{\cal K}}(\Phi^m) 
{\cal Q}^{\cal I} {\cal Q}^{\cal J} {\cal Q}^{\cal K} 
\nonumber \\ 
&=& \lambda^{(U)}_{IJK}(\Phi^m) 
Q^I U^J H_u^K 
+\lambda^{(D)}_{IJK}(\Phi^m) 
Q^I D^J H_d^K 
\nonumber \\ &&
+ \lambda^{(N)}_{IJK}(\Phi^m) 
L^I N^J H_u^K 
+ \lambda^{(E)}_{IJK}(\Phi^m) 
L^I E^J H_d^K, 
\nonumber \\*[5pt] 
f_a &=& S \quad (a = 1, 2, 3), 
\nonumber
\end{eqnarray}
respectively, where 
${\cal Q}^{\cal I}$ and $\Phi^m$ symbolically represents 
the MSSM matter and the moduli chiral superfields as shown 
in Eq.~(\ref{eq:matter}), 
$I,J=1,2,3$ and $K,L=1,2,\ldots,6$ label generations, 
and traces of the YM-indices are implicit. 

In the K\"ahler potential, the YM-field independent part 
$K^{(0)}(\bar\Phi^{\bar{m}},\Phi^m)$ is given by 
\begin{eqnarray}
K^{(0)}(\bar\Phi^{\bar{m}},\Phi^m) &=& 
-\ln (S+\bar{S}) 
-\sum_r \ln(T_r + \bar{T}_r)  
-\sum_r \ln(U_r + \bar{U}_r), 
\nonumber
\end{eqnarray}
and the K\"ahler metrics of chiral matters 
as functions of moduli are found as 
\begin{eqnarray}
Z^{({\cal Q}_L)}_{\bar{I}J}(\bar\Phi^{\bar{m}},\Phi^m) 
&=& \delta_{\bar{I}J}\frac{1}{\sqrt{3}} (T_2+\bar{T}_2)^{-1} 
(U_1+\bar{U}_1)^{-1/2} (U_2+\bar{U}_2)^{-1/2} 
\exp \frac{4 \pi \left( {\rm Im}\,\zeta_{{\cal Q}_L} 
\right)^2}{3(U_1+\bar{U}_1)}, 
\nonumber \\
Z^{({\cal Q}_R)}_{\bar{I}J}(\bar\Phi^{\bar{m}},\Phi^m) 
&=& \delta_{\bar{I}J}\frac{1}{\sqrt{3}} (T_3+\bar{T}_3)^{-1} 
(U_1+\bar{U}_1)^{-1/2} (U_3+\bar{U}_3)^{-1/2} 
\exp \frac{4 \pi \left( {\rm Im}\,\zeta_{{\cal Q}_R} 
\right)^2}{3(U_1+\bar{U}_1)}, 
\nonumber \\
Z^{({\cal Q}_H)}_{\bar{I}J}(\bar\Phi^{\bar{m}},\Phi^m) 
&=&\delta_{\bar{I}J} \sqrt{6} (T_1+\bar{T}_1)^{-1} 
\left\{ \prod_{r=1}^3 (U_r+\bar{U}_r)^{-1/2} \right\} 
\exp \frac{-4 \pi \left( {\rm Im}\,\zeta_{{\cal Q}_H} 
\right)^2}{6(U_1+\bar{U}_1)}, 
\nonumber
\end{eqnarray}
where 
${\cal Q}_L = \{ Q, L \}$, 
${\cal Q}_R = \{ U, D, N, E \}$ and 
${\cal Q}_H = \{ H_u, H_d \}$, 
and the Wilson-line parameters 
$\zeta_{{\cal Q}_L}$, 
$\zeta_{{\cal Q}_R}$ and $\zeta_{{\cal Q}_H}$ 
are defined in Eq.~(\ref{eq:dzeta}). 

On the other hand, in the superpotential, 
the holomorphic Yukawa couplings of chiral matters 
as functions of moduli are given by 
\begin{eqnarray}
\lambda^{({\cal Q}_y)}_{IJK}(\Phi^m) 
&=& \sum_{m=1}^6 \delta_{I+J+3(m-1),\,K}\,
\vartheta \left[ 
\begin{array}{c}
\frac{3(I-J)+9(m-1)}{54} \\ 0 
\end{array} \right] 
\left( 3 \left( 
\bar\zeta_{{\cal Q}_L} 
- \bar\zeta_{{\cal Q}_R} 
\right),\, 54iU_1 \right), 
\nonumber
\end{eqnarray}
where the superscript ${\cal Q}_y$ represents 
${\cal Q}_y = U, D, N, E$ 
and these also indicate corresponding subscripts 
${\cal Q}_L = Q, Q, L, L$ and 
${\cal Q}_R = U, D, N, E$, respectively, and 
$\vartheta$ represents the Jacobi theta-function: 
\begin{eqnarray}
\vartheta \left[ \begin{array}{c} 
a \\ b \end{array} \right] 
\left( \nu, \tau \right) &=& 
\sum_{l \in {\bf Z}} 
e^{\pi i \left( a + l \right)^2 \tau}
e^{2 \pi i \left( a + l \right) 
\left( \nu + b \right)}. 
\nonumber
\end{eqnarray}


\begin{thebibliography}{99}

\bibitem{Ishimori:2010au}
  H.~Ishimori, T.~Kobayashi, H.~Ohki, Y.~Shimizu, H.~Okada and M.~Tanimoto,
  Prog.\ Theor.\ Phys.\ Suppl.\  {\bf 183} (2010) 1
  [arXiv:1003.3552 [hep-th]]; 
Lect.\ Notes Phys.\  {\bf 858}, 1 (2012).  

\bibitem{ArkaniHamed:1999dc}
  N.~Arkani-Hamed and M.~Schmaltz,
  Phys.\ Rev.\ D {\bf 61} (2000) 033005  [hep-ph/9903417].

\bibitem{Abe:2009vi}
  H.~Abe, K.~-S.~Choi, T.~Kobayashi and H.~Ohki,
  Nucl.\ Phys.\ B {\bf 820} (2009) 317
  [arXiv:0904.2631 [hep-ph]].


\bibitem{Abe:2010iv}
  H.~Abe, K.~-S.~Choi, T.~Kobayashi, H.~Ohki and M.~Sakai,
  Int.\ J.\ Mod.\ Phys.\ A {\bf 26} (2011) 4067
  [arXiv:1009.5284 [hep-th]].

\bibitem{Kobayashi:2004ya} 
  T.~Kobayashi, S.~Raby and R.~-J.~Zhang,
  Nucl.\ Phys.\ B {\bf 704}, 3 (2005)  [hep-ph/0409098]; 
%
  T.~Kobayashi, H.~P.~Nilles, F.~Ploger, S.~Raby and M.~Ratz,
Nucl.\ Phys.\ B {\bf 768}, 135 (2007)  [hep-ph/0611020]; 
%
  P.~Ko, T.~Kobayashi, J.~-h.~Park and S.~Raby,
Phys.\ Rev.\ D {\bf 76}, 035005 (2007)  
[Erratum-ibid.\ D {\bf 76}, 059901 (2007)]  [arXiv:0704.2807
[hep-ph]].  

\bibitem{Angelantonj:2000hi} 
  C.~Angelantonj, I.~Antoniadis, E.~Dudas and A.~Sagnotti,
  Phys.\ Lett.\ B {\bf 489}, 223 (2000)
  [hep-th/0007090].

\bibitem{Cremades:2004wa}
  D.~Cremades, L.~E.~Ibanez and F.~Marchesano,
  JHEP {\bf 0405} (2004) 079  [hep-th/0404229].

\bibitem{Abe:2008sx}
  H.~Abe, K.~-S.~Choi, T.~Kobayashi and H.~Ohki,
  Nucl.\ Phys.\ B {\bf 814} (2009) 265
  [arXiv:0812.3534 [hep-th]].


\bibitem{Abe:2009dr} 
  H.~Abe, K.~-S.~Choi, T.~Kobayashi and H.~Ohki,
JHEP {\bf 0906}, 080 (2009)  [arXiv:0903.3800 [hep-th]].  





%
\bibitem{Abe:2009uz}
   H.~Abe, K.~-S.~Choi, T.~Kobayashi and H.~Ohki,
  Phys.\ Rev.\ D {\bf 80} (2009) 126006
  [arXiv:0907.5274 [hep-th]]; 
%
  Phys.\ Rev.\ D {\bf 81} (2010) 126003
  [arXiv:1001.1788 [hep-th]].


\bibitem{BerasaluceGonzalez:2012vb} 
  M.~Berasaluce-Gonzalez, P.~G.~Camara, F.~Marchesano, D.~Regalado and A.~M.~Uranga,
JHEP {\bf 1209}, 059 (2012)  [arXiv:1206.2383 [hep-th]].  



\bibitem{Abe:2012ya}
  H.~Abe, T.~Kobayashi, H.~Ohki and K.~Sumita,
  Nucl.\ Phys.\ B {\bf 863} (2012) 1
  [arXiv:1204.5327 [hep-th]].

\bibitem{Marcus:1983wb}
  N.~Marcus, A.~Sagnotti and W.~Siegel,
  Nucl.\ Phys.\ B {\bf 224} (1983) 159; 
%
  N.~Arkani-Hamed, T.~Gregoire and J.~G.~Wacker,
  JHEP {\bf 0203} (2002) 055  [hep-th/0101233].


\bibitem{Choi:2009pv} 
  K.~-S.~Choi, T.~Kobayashi, R.~Maruyama, M.~Murata, Y.~Nakai, H.~Ohki and M.~Sakai,
Eur.\ Phys.\ J.\ C {\bf 67}, 273 (2010)  [arXiv:0908.0395 [hep-ph]].  
%
%
  T.~Kobayashi, R.~Maruyama, M.~Murata, H.~Ohki and M.~Sakai,
JHEP {\bf 1005}, 050 (2010)  [arXiv:1002.2828 [hep-ph]].  


\bibitem{Abe:2008fi}
  H.~Abe, T.~Kobayashi and H.~Ohki,
  JHEP {\bf 0809} (2008) 043
  [arXiv:0806.4748 [hep-th]].

\bibitem{Lee:2003mc}
  H.~M.~Lee, H.~P.~Nilles and M.~Zucker,
  Nucl.\ Phys.\ B {\bf 680} (2004) 177
  [hep-th/0309195].

\bibitem{Green:1984sg}
  M.~B.~Green and J.~H.~Schwarz,
  Phys.\ Lett.\ B {\bf 149} (1984) 117.

\bibitem{Kaku:1977rk}
  M.~Kaku, P.~K.~Townsend and P.~van Nieuwenhuizen,
  Phys.\ Rev.\ Lett.\  {\bf 39} (1977) 1109.

\bibitem{Kobayashi:1973fv}
  M.~Kobayashi and T.~Maskawa,
  Prog.\ Theor.\ Phys.\  {\bf 49} (1973) 652.

\bibitem{Pontecorvo:1967fh}
  B.~Pontecorvo,
  Sov.\ Phys.\ JETP {\bf 26} (1968) 984
   [Zh.\ Eksp.\ Teor.\ Fiz.\  {\bf 53} (1967) 1717]; 
%
  Z.~Maki, M.~Nakagawa and S.~Sakata,
  Prog.\ Theor.\ Phys.\  {\bf 28} (1962) 870.

\bibitem{Beringer:1900zz}
  J.~Beringer {\it et al.}  [Particle Data Group Collaboration],
  Phys.\ Rev.\ D {\bf 86} (2012) 010001.

\bibitem{Ibanez:2006da}
  L.~E.~Ibanez and A.~M.~Uranga,
  JHEP {\bf 0703} (2007) 052
  [hep-th/0609213]; 
%
  M.~Cvetic, R.~Richter and T.~Weigand,
  Phys.\ Rev.\ D {\bf 76} (2007) 086002
  [hep-th/0703028].


\bibitem{Kaplunovsky:1993rd}
 V.~S.~Kaplunovsky and J.~Louis,
 Phys.\ Lett.\  B {\bf 306}, 269 (1993)
 [arXiv:hep-th/9303040];
%
%
%
 A.~Brignole, L.~E.~Ibanez and C.~Munoz,
 Nucl.\ Phys.\  B {\bf 422}, 125 (1994)
 [Erratum-ibid.\  B {\bf 436}, 747 (1995)]
 [arXiv:hep-ph/9308271];
%
 T.~Kobayashi, D.~Suematsu, K.~Yamada and Y.~Yamagishi,
 Phys.\ Lett.\  B {\bf 348}, 402 (1995)
 [arXiv:hep-ph/9408322];
%
 L.~E.~Ibanez, C.~Munoz and S.~Rigolin,
 Nucl.\ Phys.\  B {\bf 553}, 43 (1999)
 [arXiv:hep-ph/9812397].





\bibitem{Randall:1998uk}
 L.~Randall and R.~Sundrum,
 Nucl.\ Phys.\  B {\bf 557}, 79 (1999)
 [arXiv:hep-th/9810155];
%
 G.~F.~Giudice, M.~A.~Luty, H.~Murayama and R.~Rattazzi,
 JHEP {\bf 9812}, 027 (1998)
 [arXiv:hep-ph/9810442].






\bibitem{Choi:2005ge}
  K.~Choi, A.~Falkowski, H.~P.~Nilles and M.~Olechowski,
  Nucl.\ Phys.\ B {\bf 718} (2005) 113
  [hep-th/0503216].





\bibitem{Lust:2004cx} 
  D.~Lust, P.~Mayr, R.~Richter and S.~Stieberger,
  Nucl.\ Phys.\ B {\bf 696}, 205 (2004)
  [hep-th/0404134].

\bibitem{:2012rz}
  G.~Aad {\it et al.}  [ATLAS Collaboration],
  arXiv:1208.0949 [hep-ex].

\bibitem{Misiak:1997ei}
  M.~Misiak, S.~Pokorski and J.~Rosiek,
  Adv.\ Ser.\ Direct.\ High Energy Phys.\  {\bf 15} (1998) 795
  [hep-ph/9703442].

\bibitem{Choi:2005uz}
  K.~Choi, K.~S.~Jeong and K.~i.~Okumura,
  JHEP {\bf 0509} (2005) 039
  [hep-ph/0504037]; 
  M.~Endo, M.~Yamaguchi and K.~Yoshioka,
  Phys.\ Rev.\ D {\bf 72} (2005) 015004
  [hep-ph/0504036].

\bibitem{Choi:2005hd}
  K.~Choi, K.~S.~Jeong, T.~Kobayashi and K.~i.~Okumura,
  Phys.\ Lett.\ B {\bf 633} (2006) 355
  [hep-ph/0508029]; 
  R.~Kitano and Y.~Nomura,
  Phys.\ Lett.\ B {\bf 631} (2005) 58
  [hep-ph/0509039]; 
  K.~Choi, K.~S.~Jeong, T.~Kobayashi and K.~i.~Okumura,
  Phys.\ Rev.\ D {\bf 75} (2007) 095012
  [hep-ph/0612258].

\bibitem{:2012gk}
  G.~Aad {\it et al.}  [ATLAS Collaboration],
  Phys.\ Lett.\ B
  [arXiv:1207.7214 [hep-ex]].
%
  S.~Chatrchyan {\it et al.}  [CMS Collaboration],
  Phys.\ Lett.\ B
  [arXiv:1207.7235 [hep-ex]].

\bibitem{Ellwanger:2009dp}
 U.~Ellwanger, C.~Hugonie and A.~M.~Teixeira,
 Phys.\ Rept.\  {\bf 496}, 1 (2010)
 [arXiv:0910.1785 [hep-ph]].

\bibitem{Asano:2012sv}
  M.~Asano and T.~Higaki,
  arXiv:1204.0508 [hep-ph];
  T.~Kobayashi, H.~Makino, K.~-i.~Okumura, T.~Shimomura and T.~Takahashi,
  arXiv:1204.3561 [hep-ph].


\bibitem{DiVecchia:2011mf}
  P.~Di Vecchia, R.~Marotta, I.~Pesando and F.~Pezzella,
  J.\ Phys.\ A A {\bf 44} (2011) 245401
  [arXiv:1101.0120 [hep-th]].

\bibitem{Abe:2007kf}
  H.~Abe, T.~Kobayashi and Y.~Omura,
  Phys.\ Rev.\ D {\bf 76} (2007) 015002
  [hep-ph/0703044 [hep-ph]]; 
%
%
  S.~Antusch, L.~Calibbi, V.~Maurer, M.~Monaco and M.~Spinrath,
arXiv:1207.7236 [hep-ph].  
%
  H.~Abe, J.~Kawamura and H.~Otsuka,
  arXiv:1208.5328 [hep-ph].

\bibitem{WIP}
  H.~Abe, T.~Horie and K.~Sumita, work in progress. 

\bibitem{Kachru:2003aw}
  S.~Kachru, R.~Kallosh, A.~D.~Linde and S.~P.~Trivedi,
  Phys.\ Rev.\ D {\bf 68} (2003) 046005
  [hep-th/0301240].

\bibitem{Choi:2004sx}
  K.~Choi, A.~Falkowski, H.~P.~Nilles, M.~Olechowski and S.~Pokorski,
  JHEP {\bf 0411} (2004) 076  [hep-th/0411066].

\bibitem{Dudas:2006gr}
  E.~Dudas, C.~Papineau and S.~Pokorski,
  JHEP {\bf 0702} (2007) 028
  [hep-th/0610297]; 
%
  H.~Abe, T.~Higaki, T.~Kobayashi and Y.~Omura,
  Phys.\ Rev.\ D {\bf 75} (2007) 025019
  [hep-th/0611024];
%
  H.~Abe, T.~Higaki and T.~Kobayashi,
Phys.\ Rev.\ D {\bf 76}, 105003 (2007)  [arXiv:0707.2671 [hep-th]].  


\bibitem{Intriligator:2006dd}
  K.~A.~Intriligator, N.~Seiberg and D.~Shih,
  JHEP {\bf 0604} (2006) 021
  [hep-th/0602239].

\bibitem{Hamada:2012wj} 
  Y.~Hamada and T.~Kobayashi,
arXiv:1207.6867 [hep-th].  

\end{thebibliography}
\end{document}